\documentstyle[12pt,epsfig]{article}
\newfont{\Bbb}{msbm10 scaled 1200}     
\newcommand{\mathbb}[1]{\mbox{\Bbb #1}}

\def\TL{\hfil$\displaystyle{##}$}
\def\TR{$\displaystyle{{}##}$\hfil}

\def\lbldef#1#2{\expandafter\gdef\csname #1\endcsname {#2}}
\def\eqn#1#2{\lbldef{#1}{(\ref{#1})}%
\begin{equation} #2 \label{#1} \end{equation}}
\def\eqalign#1{\vcenter{\openup1\jot
    \halign{\strut\span\TL & \span\TR\cr #1 \cr
   }}}

\def\href#1#2{#2}
\def\half{{1 \over 2}}

\newcommand{\beq}{\begin{equation}}
\newcommand{\eeq}{\end{equation}}
\newcommand{\ber}{\begin{eqnarray}}
\newcommand{\eer}{\end{eqnarray}}

\newcommand{\beqar}{\begin{eqnarray}}

\newcommand{\eeqar}{\end{eqnarray}}

\newcommand{\ba}{\begin{eqnarray}}
\newcommand{\ea}{\end{eqnarray}}

\newcommand{\dsl}
  {\kern.06em\hbox{\raise.15ex\hbox{$/$}\kern-.56em\hbox{$\partial$}}}

\newcommand{\D}{{\cal{D}}}

\newcommand{\eeqarr}{\end{eqnarray}}
\newcommand{\ZZ}{{\rm \kern 0.275em Z \kern -0.92em Z}\;}

\begin{document}
\baselineskip=15.5pt
\pagestyle{plain}
\setcounter{page}{1}
\begin{titlepage}

\leftline{\tt hep-th/0007018}

\vskip -.8cm

\rightline{\small{\tt HUTP-99/A027}}

\begin{center}

\vskip 1.7 cm

{\LARGE {Supergravity description of field theories \\ 
on curved manifolds and a no go theorem }}

\vskip 1.5cm
{\large 
Juan Maldacena  and Carlos Nu\~nez
}
\vskip 1.2cm

Jefferson  Laboratory of Physics,
Harvard University, Cambridge, MA  02138, USA

\vskip 0.5cm

\vspace{1cm}

{\bf Abstract}

\end{center}
In the first part of this paper we 
 find  supergravity solutions corresponding to branes 
on worldvolumes of the form $R^d \times \Sigma$ where $\Sigma$ is
a  Riemann surface. These theories
arise when we wrap branes on holomorphic Riemann surfaces 
 inside $K3$ or CY manifolds.
In some  cases the theory at low energies is a conformal field theory
with two less dimensions. 
We find some  non-singular supersymmetric 
 compactifications of M-theory down  to $AdS_5$.
We also propose  a criterion for permissible 
singularities in 
supergravity solutions. 

In the second part of this paper, which can be read independently of  the
first,
we show that there are no non-singular Randall-Sundrum
or de-Sitter compactifications for  large class of gravity  
theories.

\noindent

\end{titlepage}

\newpage


\section{Introduction}
\label{intro}

In the first part of this 
 paper we  study the large $N$ limit of branes wrapped on non-trivial 
cycles. More precisely we consider a $d+2 $ dimensional field theory 
defined on a space of the form $R^d \times \Sigma_{g}$ where $\Sigma_g$ 
is a genus $g$ Riemann
surface. These theories reduce to $d$ dimensional field theories at low 
energies, at energies small compared to the inverse size of the Riemann
surface. 
We consider supergravity solutions that describe the flow between the
$d+2$ dimensional field theory and the $d$ dimensional field theory.
Our $d+2$ dimensional field theories are superconformal and 
have the maximum amount of supersymmetry. More precisely we 
consider ${\cal N} =4$ super-Yang-Mills in four dimensions and 
the $(0,2)$ theory in six dimensions. These will give rise to 
two and four dimensional field theories respectively. 
These field theories will have less supersymmetry. 
We consider situations where the supersymmetry gets reduced to 1/2 or 
1/4 of the original supersymmetry of the $d+2$ dimensional theory. 
The amount of supersymmetry preserved depends on how the two dimensional
surface $\Sigma_g$ is embedded in a higher dimensional space. 
In the field theory limit, these different embedding possibilities 
translate into different normal bundles and therefore different 
external $SO(n)$ gauge fields on the worldvolume theory, where $n$
is the number of directions normal to the brane. 
There are several cases where the resulting $d$ dimensional field
theory is conformal. In those cases we find an $AdS_{d+1}$ geometry 
in the IR. These geometries in the IR have an interest of its own
and provide new AdS/CFT examples. In particular, we find new 
${\cal N}=2 $ and ${\cal N}=1$ superconformal field theories
in four dimensions that arise from wrapping M5 branes on 
negatively curved Riemann surfaces (surfaces with $ g>1$). 
They are dual to $AdS_5$ warped compactifications of eleven dimensional 
supergravity. These are $AdS_5$ fibrations over a six dimensional 
space, the five dimensional warp factor depends on the six dimensional
coordinates\footnote{Related compactifications were found in 
\cite{Alishahiha:1999ds,Cvetic:2000cj}, but 
in their case they the warp factor was going to zero
at some points in the internal space. In our case the 
metric is completely smooth. In \cite{Fayyazuddin:1999zu} similar 
configurations were studied and the conditions for preserved supersymmetry
were studied in great generality and some solutions were found
corresponding
to intersecting branes with a certain amount of ``smearing'' in one of
 the
transverse
dimensions.}.
Similarly we find some compactifications of IIB string theory 
to $AdS_3$ that  preserve (2,2) supersymmetry. 

The basic technique that we use to find the solutions is the following. 
The field theories on branes are twisted theories 
\cite{Bershadsky:1996qy}. What 
this means is that together with the coupling to the curvature
of the brane worldvolume there is a coupling to an  external 
$SO(n)$ gauge field, if $n$ is the number of transverse directions
to the brane. So we have a field theory on a curved space coupled
to a $SO(n)$ gauge field.
In supergravity this translates into
boundary conditions at the boundary of $AdS_{d+3}$ for  the metric
and the $SO(n)$ gauge fields. Fortunately these are modes of gauged
supergravity, so that we can consider just the $d+3$ dimensional 
gauged supergravity
equations. We further consider only constant curvature Riemann surfaces
and  simple embedding of the spin connection into  the $SO(n)$
connection. A more general analysis is left for the future.

Some of these supergravity solutions can be viewed as compactifications
of $d+3$ dimensional gauged supergravities on $AdS_{d+1} \times
\Sigma_{g>1}$ with magnetic fluxes on $\Sigma$. They are similar in 
spirit to the solutions in \cite{Freedman:1984rs,Chamseddine:1999xk,%
Klemm:2000nj,Cacciatori:2000bw,Chamseddine:2000bk}.

Since we have obtained a large family of $AdS_5$ compactifications 
it is natural to ask if we could find a smooth  Randall-Sundrum 
\cite{Randall:1999vf} compactification
based on these.  In  section \ref{nogo} 
we show  that it is not possible
to find a smooth Randall-Sundrum compactification of usual 
supergravity theories. 
  In fact we prove something even more general, it is 
not possible to find warped compactifications which have only 
singularities where the warp factor is non-increasing as we approach the 
singularity. We also show that
there are no deSitter compactifications. This section is self contained
and can be read independently of the rest of this paper.
For other no go theorems for supersymmetric RS compactifications of 
5d gauged supergravity
see \cite{Kallosh:2000tj,Wijnholt:1999vk,Gibbons:2000hg}.

There are many papers in the literature which consider compactifications
similar to the ones considered here 
\cite{Papadopoulos:1999tw,Chamseddine:1999xk,%
Klemm:2000nj,Cacciatori:2000bw,Chamseddine:2000bk,Yang:1999ze,Kiem:1998ga,%
Youm:1999zs, Loewy:1999mn,Gomberoff:1999ps,Youm:1999tt,Cherkis:1999jt}.
While this paper was in preparation the paper \cite{Fayyazuddin:2000em}
appeared
which  has some overlap with ours and presents another interesting
method, based on \cite{Fayyazuddin:1999zu},
for obtaining a large class of solutions corresponding
to branes on Riemann surfaces.

This paper is organized as follows. We first describe the general 
idea behind twisted field theories and how we plan to find the 
gravity solutions. In section 3 we consider ${\cal N} =4$ super Yang
Mills on $R^2 \times \Sigma$. We consider two possible twisting 
which preserve $(4,4)$ and $(2,2)$ supersymmetry in 1+1 dimensions. 
In section four we consider the (0,2) superconformal field theory 
that lives on M5 branes compactified on spaces of the form 
$R^4 \times \Sigma$. We consider again two cases, preserving 
${\cal N} = 2,1$ supersymmetry in four dimensions.
In section \ref{criterion}
 we discuss a criterion for allowed singularities in gravity
theories that are dual to field theories. 
In section \ref{nogo} we discuss 
the absence of certain  warped compactifications
or
de Sitter solutions in 
a class of gravity theories, including 11d supergravity, IIB, IIA and
massive IIA.

In the appendix  we give some more details on our calculations.

\section{General Idea}

If we start with a supersymmetric field theory and we put it on a 
curved manifold $\Omega$  then, in general, we will break supersymmetry
since we will not have a covariantly constant spinor, obeying
$(\partial_\mu + \omega_\mu ) \epsilon =0 $.
If the supersymmetric theory has a global R-symmetry, then 
we can couple the theory to an external gauge field that couples
to the R-symmetry current. If we choose the external gauge 
field to be equal to the spin connection $A_\mu = \omega_\mu$ (the
precise meaning of this equation will be explained below)  then we 
see that we can find a covariantly constant spinor since
$\partial_\mu + \omega_\mu - A_\mu \epsilon = \partial_\mu \epsilon $,
so that we  can just consider a constant spinor. 
The resulting theory is a so called ``twisted'' theory, since we can
view the  the coupling
to the external gauge field  as effectively changing
 the spins of all fields. The supersymmetry parameter becomes a scalar. 
Though it  might sound like a contrived way of preserving supersymmetry,
it  is precisely the way that branes wrapping on non-trivial cycles
in M-theory or string theory compactifications 
manage to preserve 
some supersymmetries \cite{Bershadsky:1996qy}. In
this case $\Omega$ is the worldvolume 
geometry of the cycle and the external gauge field takes into account
 the 
fact that the directions normal to the cycle form a non-trivial bundle, 
the normal bundle, and $A_\mu$ is the connection on this normal bundle. 
The condition that the cycle preserves some supersymmetry then boils
down to the condition that the spin connection is equal to the gauge 
connection. 
So if we are interested in understanding field theories arising 
on branes wrapping non-trivial cycles, then we will have to study these
twisted theories. 
Let us first clarify the nature of the limit in which we decouple
this twisted field theory from the full original string theory. 
We consider a  brane wrapping
a cycle and we take the decoupling limit $l_p\to 0$ as in 
\cite{Maldacena:1998re}
keeping the volume of the cycle fixed. In this limit we get a field 
theory on the brane that is twisted as above, due to the non-trivial
embedding of the cycle in the ambient space. If the theory on the 
brane is conformal before we wrap it, this is all we have 
to do. If we have a D-p-brane with $p \not = 3$ then we also should 
scale the string coupling as in the flat case  \cite{Itzhaki:1998dd}. 
Notice that in this limit the theory is not sensitive to the global
geometry of the spacetime where the cycle is embedded. The reason 
is that a finite fluctuation of the scalar field parameterizing 
transverse displacements corresponds to an infinitesimal fluctuation
in the position of the brane. In other words, in the scaling limit
we are considering, the size of the Calabi Yau (or any other space where the 
brane is embedded) is fixed, while the 
typical fluctuation of the position of the 
brane goes to zero as  $l_p$ goes to zero. 
This of course does not imply that the scalars associated to motions
of the brane 
will have definite expectation values. Whether they do or not will 
depend on the number of non-compact dimensions of the theory. We will
discuss this more later.
Let us consider some examples. Suppose we have type IIB string theory 
on $R^6 \times K3$. We can wrap   a D3 brane 
 on an $S^2$ inside the  $K3$  manifold leaving two non-compact 
directions. 
The worldvolume is then $\Omega = R^2 \times S^2$. In this case,
the spin connection
is in a $U(1)$ subgroup of the tangent group $SO(3,1)$,
 since the curvature is purely in
the $S^2$ directions. The brane has 6 normal directions. Two of them
are in the directions of $K3$ that are normal to the $S^2$. 
They will form a non-trivial $U(1)$ normal bundle.
It turns out that for holomorphically embedded spheres, the spin connection
is essentially  equal to the connection on the $U(1)$ normal bundle. 
The other four normal directions are totally flat, with no gauge field. 
This gauge connection will break the $SO(6)$ R-symmetry group into
$U(1) \times SO(4)$. Out of the 16 spinors that generate the 
supersymmetries of ${\cal N}=4$ Yang Mills, there will be only half
for which the chirality on $S^2$ and the $U(1)$ charges are correlated 
so that the spin connection and gauge connection cancel. In terms of
the two dimensional low energy theory on $R^2$ the theory has $(4,4)$ 
supersymmetry. These kind of theories were considered in 
\cite{Vafa:1994tf,Bershadsky:1995vm}. 
Below we find the supergravity solutions associated to these twisted 
field theories. For this particular example we find that there is 
a family of supergravity solutions and that they all seem to  have a 
singularity in the IR. We argue later that this singularity is associated
to the  IR properties of the brane theory and that only the 
singularities of an allowable type, according to the criterion
in section \ref{criterion}  or the one in  \cite{Gubser:2000nd}, 
produce the right physics.  
We will consider however some other examples, involving
branes wrapping negatively curved Riemann surfaces, for which there
are non-singular solutions which in the IR have an $AdS$ form. 
For example if we wrap a $D3$ brane on a genus $g >1$ Riemann surface
times $R^2$ with a particular normal bundle we specify below then
 the supergravity solution  interpolates between 
an $AdS_5$ region close to the boundary 
and an $AdS_3$ region in the IR, corresponding
to the fact that the theory in 
the UV is
just $3+1 $ dimensional SYM and it 
 flows to a 1+1 dimensional conformal field
theory. 
So these are examples of flows ``across dimensions'', the four
dimensional
conformal field theory flows to a two dimensional conformal field theory.

We only consider cases where the genus of the  Riemann surface $\Sigma_g$
is 
 $g\not =1$, since in the case of 
$g=1$ the constant curvature metric is just the metric on a flat $T^2$
and the theory is identical to the untwisted theory.

We will find similar examples for $M5$ branes wrapping on negatively 
curved Riemann surfaces. These will give us a  new family of 
examples of four dimensional conformal field theories and their
associated $AdS_5$ compactifications of M-theory. The case analyzed
in  \cite{Alishahiha:1999ds,Oz:1999qd}
 would probably  arise as a singular limit of the smooth 
solutions analyzed here. In principle it should also be 
 possible to 
find the solutions in our paper using the methods of 
\cite{Fayyazuddin:1999zu,Fayyazuddin:2000em} but we will not pursue that
here.  
In Horava-Witten M-theory compactifications to 3+1 dimensions
we could have 
M5 branes wrapping on Riemann surfaces in the CY manifold at some points
in moduli space.
Our analysis implies  that
coincident branes will typically give rise to conformal field
theories. 
More accurately, that they give rise to conformal field theories
in the large $N$ limit and the large CY volume limit. Finite $N$ effects,
or finite volume effects of the CY manifold could destroy conformality. 
Since having a conformal field theory, or a theory with a logarithmically
running coupling would give a natural explanation of the gauge 
hierarchy problem, it is  interesting that these arise quite 
naturally in these compactifications. 
One should also be careful with this
conclusion because in 
 Horava-Witten compactifications we will also generically have 
some fluxes of the four form field strength. These might induce relevant
perturbations of the field theory, as in \cite{Polchinski:2000uf}. 

Since we are dealing with the theory of coincident M5 branes we find it
hard to give a purely four dimensional field theory description
of the theory in the IR. 
 It looks like it should be possible to say
what it is  more precisely, but leave this to the future.
Dimensional reducing the M5 solutions we get D4 solutions corresponding
to D4 branes wrapped on Riemann surfaces. In this case we can state
more precisely what the 3 dimensional field theories are.

In this paper we will consider Riemann surfaces with constant curvature.
We think of them as $H_2/\Gamma$ where $\Gamma$ is a discrete subgroup
of $SL(2,R)$. 
Notice that for our purposes, which is to find a precise sugra solution,
the precise metric on the Riemann surface 
 {\it does } matter. 
One can find solutions for non-constant curvatures by the methods of
this paper or by the methods of
 \cite{Fayyazuddin:2000em} where a large family of solutions was found.

\section{Twisted 4d N=4 SYM}

The possible twistings of $N=4$ SYM were considered in \cite{Vafa:1994tf}. 
These twistings are distinguished by different ways of embedding 
the spin connection in the $SU(4)$ R-symmetry group. 
In principle we could consider the most general case by the techniques
of this paper. But in order to keep formulas simple we will 
concentrate on four dimensional spaces of the form $R^2 \times \Sigma_2$
where
$\Sigma_2$ is a two dimensional manifold. In this case the spin connection
is in a U(1) subgroup. So, different twistings of the theory are
defined by different embedding of this $U(1)$ in $SU(4) =SO(6)$. 
We will consider two cases, one which preserves (4,4) supersymmetry and
one preserving (2,2) supersymmetry. 

Throughout this section we work in units where the five dimensional 
$AdS_5$ radius is one. In order to restore the radius dependence of
the solutions we just multiply the five or ten dimensional metrics
we will write  by $R_{AdS_5}^2   = \sqrt{ 4 \pi g_s N } \alpha' $.

\subsection{Twists preserving $(4,4)$ susy}

The first twisting that we consider 
corresponds to picking a $U(1) $ in SO(6) in such
a way that we break $SO(6) \to SO(2) \times SO(4)$, where the first
$SO(2)$ is the $U(1)$ that we are picking inside $SO(6)$. 
In other words, if we think of $SO(6) $ as acting on six coordinates
$\phi^I$, then the $U(1)$ is the group of rotations in the 12 plane. 
This twisting turns out to be exactly the same  as the one  considered
in \cite{Vafa:1994tf}\footnote{
In \cite{Vafa:1994tf} more general manifolds were considered, but it 
was noted that if the four dimensional manifold is Kahler (as in our
case),
 then 
only the $U(1)$ mentioned above has a non-trivial gauge connection}
and the resulting field  theories were analyzed in \cite{Bershadsky:1995vm}.
 In order to explain more clearly how this works, 
let us  consider a  field $\phi$ in the Yang-Mills Lagrangian, 
with spin $s$ under the $SO(2)$
 spin connection on $\Sigma_2$ and $U(1)$ charge
$q$. The Lagrangian is obtained from the flat Lagrangian, by replacing
ordinary derivatives of $\phi$ in the $\Sigma$ directions  
 by covariant derivatives
\eqn{cov}{
{\cal D}_\mu \phi  = (\partial_\mu + s i  \omega_\mu + i q  A_\mu )\phi 
}
where $\omega_\mu = \epsilon_{a b} \omega_{\mu}^{ab}/2 $ with 
$\omega_\mu^{ab}$ the usual spin connection. 
If the metric on $\Sigma_2$ is $ds^2 = e^{2 h} (dx^2 + dy^2)$ then
the spin connection is 
$\omega_\mu = \epsilon_{\mu \nu} \partial_\nu h $, and
we also have $A_\mu = \omega_\mu$. This implies that spinors with
$s = - q$ can be  covariantly constant. In fact they are actually 
constant, the twisting effectively made them scalars.  
In this situation the  $SO(1,3) \times SO(6)$ symmetry group of the
 tangent and normal bundles  is 
decomposed as  $SO(1,1)\times SO(2)_\Sigma  \times  U(1)  
 \times SU(2)_L \times SU(2)_R $ and
the preserved spinors transform in the $(+,\pm,\mp , 1,2)$ and 
$(-, \pm, \mp ,2,1)$ representations. 
So we have (4,4) supersymmetry,  in $1+1$ dimensional notation.
Let us consider some examples where these theories arise.
Suppose we have a compactification of the form $R^6\times K3$. 
Then we can wrap a $D3$ on a holomorphic Riemann surface inside 
$K3$ and we  obtain a field theory on 
the $D3$ which is of the above from. The $SO(4) = SU(2)_L\times SU(2)_R$
symmetry of the field theory is the rotational symmetry in the four
directions in $R^6$ that are  orthogonal to the brane. 
More precisely, the limit in which we get the above field theory is a
limit where we take $\alpha' \to 0$ keeping the size of the Riemann
surface (and the $K3$) fixed. Therefore,  this limit corresponds
to a large volume compactification, large in string units. In this limit
the directions normal to the Riemann surface become effectively
 non-compact,
since the brane explores only an infinitesimal  neighborhood of the surface.
In other words, a field $\phi$ parameterizing fluctuations of the position
of the brane is related to the actual displacement by $ r = \alpha' \phi$
so that the actual displacement goes to zero as $\alpha' \to 0$. 
We can then take a further low energy limit where we consider 
energies much smaller than the size of the Riemann surface.
In this limit we obtain a two dimensional effective theory, which in the 
IR is a (4,4) superconformal theory \cite{Bershadsky:1995vm}. 
With the order of limits that we took, this conformal field theory 
has a non-compact target space since we are only exploring a neighborhood
of the Riemann surface. The fact that the target space is 
non-compact implies that this 
CFT does not have a well defined vacuum state. 
We would have obtained a different theory if
we had taken the size of the Riemann surface and K3 fixed in string units.
In that case the target space would have been essentially compact\footnote{
It still might still have some  non-compact directions via 
effects such as the ones discussed  in 
\cite{Seiberg:1999xz}.},  and quantum
 fluctuations
would have explored the whole moduli space of Riemann surfaces. The
 resulting
theory would be  (a T-dual version of) the familiar $D1-D5$ system,
if the genus of the Riemann surface is $g >1$.

Let us describe  in  more detail about the Lagrangian of these 
theories. The full  Lagrangian can be found in  \cite{Vafa:1994tf}. 
Here we give only some  terms that are relevant in 
comparing with gravity solutions. 
The only parts  of the Lagrangian that are  different from 
the usual ${\cal N}=4$ Lagrangian are  the terms
involving covariant derivatives  along $\Sigma$ or fields that are
charged under the $U(1)$ part of the normal bundle that has
a non-zero gauge connection. 
Let us consider the part of the Lagrangian involving the two 
twisted scalar fields. These are the two scalar
fields parameterizing fluctuations of the surface in
the two normal directions which transform under $U(1)$. 
Let us  arrange these two fields into
a complex field ${\cal Z} = X^1 + i X^2 $. The quadratic terms in 
the Lagrangian involving these fields are 
\eqn{cuad}{
S = \int Tr\{  |{ \cal D}_z {\cal Z} |^2 + | {\cal D}_{\bar z} {\cal Z}|^2
+{1 \over 4}  R |{ \cal Z}|^2 \}
}
The first two terms are the terms we would obviously expect from \cov\
and the last term is a curvature coupling that was derived from
supersymmetry
in 
\cite{Vafa:1994tf}, but in the case of single D-brane it  can also 
 be obtained
directly by expanding the Nambu action for a brane on this surface. 
The other scalar fields $\phi^I, ~I=1,...,4$ have a simple 
Lagrangian of the form $ (\partial \phi^I) ^2$. 
Note that integrating by parts the first term in \cuad\ we can recast
it as the second  term  up to a commutator $ [ {\cal D}_z , { \cal D}_{\bar
z} ]$. This commutator precisely cancels the curvature term in \cuad .
We can choose a holomorphic basis for the normal bundle so that
${\cal D}_{\bar z} {\cal Z} = \partial_{\bar z} {\cal Z} $.
 This implies that
holomorphic sections of the normal bundle  ${\cal Z}(z)$ are 
 solutions of 
 the 
equations of motion and describe configurations with the same energy
as the original configuration. 
This 
is precisely what we expect since any holomorphic deformation of the 
surface  preserves supersymmetry. Of course, whether these
deformations exist globally on the surface or not   depends on the global
aspects
of the geometry and the normal bundle.

Let us  look for the supergravity dual of these field 
theories. Since these are just ${\cal N}=4$ 3+1 dimensional SYM theories on
some particular backgrounds, we expect to be able to find 
the gravity dual by  starting with $AdS_5 \times S^5$ and 
changing  the asymptotic boundary 
conditions to reflect the fact that the theory is defined on $R^2\times
\Sigma$  and is coupled to an $SO(6)$ gauge field. This is easy to 
achieve. We impose that at the boundary of $AdS_5$ the metric
behaves as 
\eqn{condmet}{
ds^2 \sim  { -dt^2 + dz^2  + e^{2 h} (dx^2 + dy^2)  + dr^2 \over r^2 } 
~~~{\rm for~small}~ r
}
where $ds_\Sigma^2 = e^{2 h(x,y)} ( dx^2 + dy^2) $ is the metric of the
two dimensional surface. 
Similarly we impose that the $AdS_5$ SO(6) gauge fields 
are asymptotic to the corresponding field theory values. This translates
into a condition on components of the metric with one index in $AdS_5$ and
one index on $S^5$.  In other words we require $g_{\phi \alpha} \sim 
A_{\alpha} \sim 
\epsilon_{\alpha}^{~\beta }\partial_\beta h $ near the boundary. 
These two conditions are rather obvious. A bit less obvious is the fact
that we also need to turn on an operator in the ${\bf 20}$ of SO(6). 
This  becomes apparent once we look
at  the curvature coupling in \cuad\ and realize that that coupling
is not present for the other four scalar fields. This means that the 
operator 
\eqn{otwo}{
{\cal O}_2 =
 Tr[ {2 \over 3}  |{\cal Z}|^2  -
 {1 \over 3} ( {\phi^1}^2 + ...+ {\phi^4}^2) ]
}
 is
turned on. We also have the singlet operator turned on, with
the expected coefficient proportional to the scalar curvature 
\cite{Witten:1998qj}. So we can rewrite the curvature coupling in 
\cuad\ as
\eqn{cuadnew}{
 S  = {1 \over 2} \int   R \left( {1 \over 6} ( |{\cal Z}|^2  +
 {\phi^1}^2 + ...+ {\phi^4}^2 )  + {1 \over 2 } {\cal O}_2  \right)
} 
Fortunately all the operators that are turned on correspond
to fields in the five dimensional gauged supergravity 
multiplet. This is a general feature for these twisted theories,
even in the most general curved backgrounds. 
In order to find the gravity solutions we can therefore
consider just the five dimensional gauged supergravity equations. 
If we were interested in the most general twisted theory we would have
to use the full ${\cal N} = 8$ gauged supergravity of \cite{Gunaydin:1985qu}. 
However, 
in our  case the connection is in $U(1)$ so we can further 
use a $U(1)$ truncation of the equations of the form considered in 
\cite{Cvetic:1999xp}.
 Furthermore, in \cite{Cvetic:1999xp} one can find formulas
to express the ten dimensional solution given any solution of
the truncated five dimensional equations. 
We will consider therefore a theory involving the five dimensional
metric, a $U(1)$ gauge field and a scalar field $\varphi$ which
is dual  to the operator ${\cal O}_2$ appearing above. 
In order to find supersymmetric solutions we look at the supersymmetry
variation equations of the fermionic fields. These 
can be read of from \cite{Gunaydin:1985ak,Chamseddine:1999xk}
 as explained in the  appendix. 
\eqn{transf1}{\eqalign{
\frac{1}{\sqrt{6}}\delta \lambda &= -
\frac{1}{24} e^{-2\varphi} \Gamma^{\mu\nu}F_{\mu\nu}\epsilon
-\frac{i}{4}\Gamma^\mu
\partial_\mu\varphi \epsilon + \frac{i}{6}(e^{2\varphi}
-e^{-\varphi})\epsilon
\cr
\delta\psi_\mu &= \D_\mu(\omega)\epsilon +
\frac{i}{24}e^{-2\varphi}(\Gamma_\mu^{\nu\rho} - 4 \delta_\mu^\nu
\Gamma^\rho)F_{\nu\rho}\epsilon +\frac{1}{6}\Gamma_\mu(2 e^{-\varphi} +
e^{2\varphi})\epsilon -\frac{i}{2}A_\mu\epsilon
}}

In order to solve this equation we
 make the following ansatz for the metric
\eqn{ans}{
ds^2 = e^{2f} (dr^2 + dz^2 - dt^2) + \frac{e^{2g}}{y^2}(dx^2 + dy^2)
}
where $f$ and $g$ are functions of $r$ to be determined. 
For simplicity we are considering constant  curvature 
Riemann surfaces of genus $g>1$ \footnote{
The reader should not be confused by the fact that $g$ denotes both
the function $g(r)$ appearing in \ans\ 
 and the genus.}.
When
$r \to 0$ the boundary conditions are $f(r) \sim g(r) \sim - log(r) $.  
The gauge field is $A_x  = 1/y$. This value comes from demanding
equality
with the spin connection. 
Setting to zero the supersymmetry variations we obtain, see the appendix,
\eqn{susycond}{\eqalign{
g' &= -e^{f} [\frac{ 1}{3} (2 
e^{- \varphi }
+ e^{2\varphi}) -\frac{1}{3} e^{-2g
-2\varphi}]
\cr
f' &= -\frac{1}{6}e^f  [  
2 (2 e^{-\varphi } + e^{2 \varphi }) +  
e^{- 2g -2 \varphi}]
\cr
\varphi' & = \frac{1}{3}e^{f}[2 (- e^{-\varphi } + e^{2 \varphi })
+ e^{- 2g  -2 \varphi}]
}}

The solution is given by 
\eqn{sol}{\eqalign{
e^{- 3 \varphi}  &= 1 + e^{ - 2 \rho } 
[ {1 \over 2} \log( e^{2 \rho} - {1 \over
2} ) + C_1]
\cr
e^{ 2 g } &= e^{ 2 \rho  } e^{ -  \varphi}
\cr
e^{2 f}  & = ( e^{2 \rho } - \half ) e^{-\varphi}
\cr
({ dr \over d\rho})^2 e^{2 f} & =  { e^{ 4 \rho } \over 
(  e^{2 \rho } - \half )^2 } e^{ 2 \varphi}
}}
Notice that using the last line we can rewrite the metric in terms
of the new radial coordinate $\rho$, $\rho \to + \infty $ corresponds to 
the boundary. $C_1$ is an integration constant,
there is of course another trivial integration constant which 
amounts to shifting $f$ by a constant. 
Note that once we solve the equations for $H_2$ we can quotient the
solution by a subgroup $\Gamma$ of $SL(2,R)$ that produces the Riemann 
surface $\Sigma_g = H^2/\Gamma$. This group also acts in the $U(1)$
that we are twisting. That implies that the spinor parameter that
generates the preserved supersymmetry transforms like a scalar under
$SL(2,R)$ transformations and it therefore survives  the
quotienting procedure.

We see that there is a singularity  of the metric at $\rho=\rho^*$ where 
$e^{-3\varphi(\rho_*)}  =0$. This singularity is qualitatively the 
same regardless of the value of $C_1$.
These singularities are related to regions in the Coulomb or Higgs branches
of the theory.  
Let us explore this a bit more.
If we expand the first term  in equation  \sol\  for large $\rho$ we find
that $\varphi \sim  - e^{-2 \rho} \rho  - e^{-2 \rho} C_1 $. 
The logarithmic term
is related to the fact that we are inserting the operator ${\cal O}_2$  
as in \cuadnew\ the subleading  term can be thought of as the 
expectation value of this operator \cite{Balasubramanian:1998sn}.
We would like to relate the sign of $C_1$ to the sign of the 
expectation value of ${\cal O}_2$. We can do that by 
looking at  \cite{Freedman:1999gk} where 
 configurations in  ${\cal N}=4$ super Yang-Mills with both 
expectation values for  ${\cal O}_2$ were considered. We find from 
their paper that  $C_1 \sim  \langle {\cal O}_2 \rangle$. 
{}From the explicit expression for the operator
${\cal O}_2$ \otwo\ we see
 that configurations with $\langle {\cal O}_2 \rangle >0$ correspond
to Higgs branch configurations while ones where 
$\langle {\cal O}_2 \rangle < 0$ correspond to the Coulomb branch.
Since we also have the operator inserted, it is a bit hard
to separate its expectation value, so we should trust this criterion
only for $|C_1| $ large.
The behavior at the singularity is very similar
for both signs of $C_1$. 
These singularities look very much like the singularities in
the Coulomb branch, where branes are distributed on an $S^3$ as
in  \cite{Freedman:1999gk}, even though we expect that negative values
of $C_1$ should correspond to the Higgs branch.
This seems to be related to the Higgs-Coulomb correspondence as in 
\cite{Witten:1997yu,Berkooz:1999iz,Aharony:1999dw} where it was
shown that some singularities in the Higgs branch look very similar
to the near region of the Coulomb branch. 
What seems surprising, when we compare this system to the 
familiar D1-D5 system, is that we do not find any solution 
that has an $AdS_3$ region in the IR. 
We think that this is related to the fact that 
the (4,4) superconformal field theory that we are dealing with
here has a non-compact target space. 
This non-compactness is different in nature from the one 
that arises at special points in moduli space for the D1-D5 system
\cite{Seiberg:1999xz}. In the D1-D5 system the singularities are 
small instanton singularities which require tuning of order $Q_5$ parameters
out of the order $Q_1Q_5$ parameters in instanton moduli in order to reach
them.
In our (4,4) we have another kind of non-compactness of the moduli
space. This non-compactness is related to the possibility of 
moving the branes in the directions orthogonal to the Riemann surface.
 In other words, we are saying
that the non-compactness of the Higgs branch in this case is 
very similar to the non-compactness we would have if we 
were to consider the D1-D5 system but with the internal space replaced
by $R^4$ instead of $T^4$ and a finite number $Q_1$ of one branes.
It was shown in \cite{Marolf:1999uq} that that system has no $AdS_3$
region.

Let us find  the theory we expect at low energies from the field
theory point of view. 
 It was argued in \cite{Bershadsky:1996qy}
that this (4,4) superconformal field theory was a conformal field theory
whose target space is Hitchin space. Its central charge is $c = 6
N^2(g-1)$. This moduli space is non-compact and that seems to be preventing
the $AdS_3$ solutions. 
Another way to say what the low energy theory is; is the following.
Suppose 
we add two more dimensions along the brane so that we have a D5 brane
wrapped on $\Sigma_g$, $g>1$. Then we have an ${\cal N}=2$ theory 
in four dimensions. This four dimensional theory is a $U(N)$ 
theory with $ g $ adjoint hypers.
A way to see this is to go to weak coupling and calculate  explicitly 
the number of low energy modes and their susy properties, see
\cite{Katz:1996ht} for a related discussion. 
We can calculate that number using an  index argument. 
Supersymmetry then determines
the form of the Lagrangian.
 This theory has 
a Higgs branch of  real dimension $4 N^2(g-1)$. We can now dimensionally 
reduce two of the dimensions of the D5 brane to make it a D3 brane. 
So in the IR we have a CFT which is a sigma model whose target space
is the Higgs branch of the gauge theory. A way of thinking about a
point in this Higgs branch was given in \cite{Callan:1997kz}. The $N$
coincident
branes are moved and their intersections ``blown up'' in such a 
way that we have a single surface which now has $2 N^2(g-1) $ geometric
moduli and the same number of $U(1)$ worldvolume gauge field Wilson lines. 
Of course, in the $\alpha'\to 0 $ limit that we are taking this single 
surface has a thickness smaller than $\alpha'$ and only explores 
a small  neighborhood of the original surface but not the global structure
of the manifold where the surface $\Sigma$ is embedded.

The  singularities in \sol\ are allowed 
 under the criterion  given in section \ref{criterion}  which demands that
 $g_{00}$ does not increase as we approach the singularity 
 or the one given in 
\cite{Gubser:2000nd}, which demands that the potential for the 
scalars should be bounded above in the solution. 
See the appendix  for the explicit computation of the potential.

%
%

It is also possible to find the solution for the case that
the branes are wrapped on $R^2 \times S^2$.
Since  we can go from the metric in $H_2$, 
$ds^2= d\theta^2 + \sinh^2 \theta d\psi^2 $,  to minus the metric
on $S^2$  by taking $\theta \to i \theta$ we can find the 
equations  for the branes on $R^2\times S^2$ 
by formally replacing $e^{2 g} \to -e^{2 g} $
on the sphere by taking $\theta \to i \theta$ 
in \susycond .
Then we get a solution similar to \sol\
  that reads
\eqn{solesf}{\eqalign{
e^{- 3 \varphi}  &= 1 + e^{ - 2 \rho  } 
[ - {1 \over 2} \log( e^{ 2 \rho } + {1 \over
2} ) + C_1]
\cr
e^{ 2 g } &= e^{ 2 \rho  } e^{ -  \varphi}
\cr
e^{2 f}  & = ( e^{ 2 \rho } + \half ) e^{-\varphi}
\cr
({ dr \over d\rho })^2 e^{2 f} & =  { e^{ 4 \rho } \over 
(  e^{ 2 \rho } + \half )^2 } e^{ 2 \varphi}}}

Again  we interpret $C_1$ as giving  the expectation
value of the operator ${\cal O}_2$. 
In this case something interesting happens. 
For large enough negative  values of $C_1$ we have a singularity in 
the IR that looks very similar to that analyzed for the case of 
$H^2$. 
For  positive enough  values
of $C_1$, corresponding to positive  values of 
$\langle {\cal O}_2 \rangle$, we find that the singularity is {\it not}
allowed by the criterion in section \ref{criterion} or  Gubser's criterion 
\cite{Gubser:2000nd}. This is very 
fortunate because we do not expect this field theory to have
a Higgs branch. This is due to the fact that the scalars normal to 
the sphere have no zero modes as can be seen from the fact 
that \cuad\ is positive for non-zero values of $|{\cal Z}|$ if $R>0$.
The reduction of $N=4$ SYM on $S^2$ was 
studied in \cite{Imaanpur:1998rc}.
This theory does have a Coulomb branch
and indeed we see that in the supergravity solution.

When we talk about
Coulomb and Higgs branches  through this section
 we should remember that vacuum expectation values  for massless
fields are not well defined in two dimensions. So we should interpret
these solutions as describing some semi-classical states were some 
aspects do not change very quickly with time.
Once the singularities get resolved by the full theory, by taking 
into account the IR degrees of freedom that give rise to the singularities
then we expect that the solution would change slowly with time
as the vevs of massless fields drift through moduli space.

\subsection{ Twists preserving (2,2) supersymmetry }

Another way in which we can embed the $U(1)$ group of the spin connection
is the following. Consider the $SO(2)^3$ subgroup of $SO(6)$ which
corresponds to rotations within three orthogonal planes. Let us call these
three generators $T_i$, $i=1,2,3$. 
We 
can take the generator $T= \half  T_2 + \half T_3$.
 This breaks  $SO(6) \to
SO(2) \times SO(2) \times SU(2)$. Under this choice of subgroup the 
scalars split into two for which the normal bundle  is trivial and four
for which the normal bundle is such that they effectively become
spinors. Another way to state this choice is to consider
the splitting of $SO(6) \to SO(2) \times SO(4)$ that we considered 
in the previous section but now we take the gauge connection in 
 $U(1) \subset  SU(2)_L \subset SO(4)$.

It can be shown that a Riemann surface in a CY manifold typically 
has a normal bundle which is topologically like  the one  we are
considering
here \cite{vafacom}.
 We expect then that if we wrap a D3 brane on this
Riemann surface we  get a theory that is  similar to the theories
that we are considering here. In these more general theories the 
normal bundle does  not generically have the additional $SU(2) $ symmetry
mentioned above.

Let us consider the bosonic Lagrangian 
for this theory. Let us call 
${\cal W}_1, {\cal W}_2$ the 
two complex fields parameterizing the four directions
where the gauge field is nonzero. 
This Lagrangian will be similar to \cuad\ but the coefficient
of the curvature term will be different. 
 In principle we should get 
these terms by demanding supersymmetry. However we will find them by 
demanding holomorphicity, as in the discussion below \cuad .
The Lagrangian is then
\eqn{cuadtwo}{
S = \int Tr\{ \sum_i  |{ \cal D}_z {\cal W}_i |^2 + | {\cal D}_{\bar z}
{\cal W}_i|^2
 + {1 \over 8}  R |{ \cal W}_i|^2 \}
}
Again the remaining two untwisted scalars have an action of
the form $\int (\partial \phi)^2 $. 
The presence of the curvature term in \cuadtwo\ implies,
as in the discussion around \cuadnew , 
that we are turning on the operator in the ${\bf 20}$. It turns
out to be the same operator that we were turning on before but
with a different coefficient. 

When we consider the IR limit of this theory, at energies smaller than
the size of the Riemann surface again we expect to find a
 conformal field theory.
We can compute the central charge of this theory by an anomaly argument. 
When we reduce the theory to two dimensions we  find left and 
right moving massless fermions with different transformation properties
under the R-symmetry group. Let us classify the massless fermions 
according to their transformation properties under the 
$SO(2)\times SO(2) \times SO(2)$ subgroup of $SO(6)$. These
fermions come from the four dimensional gauginos. 
There are four possibilities for their charges:
(A) (+,+,+), (B) (-,-,+), (C) (-,+,-) (D) (+,-,-) . These 
have positive chirality in $SO(6)$, 
their complex conjugates  have negative chirality in $SO(6)$.
Now we can calculate the number of fermion zero modes on $\Sigma_g$
via an index theorem. Let us do the calculation in the weak coupling
limit where we have essentially $N^2$ free fields. These are fermions
which are coupled to an external gauge field, which is the gauge
field implementing the twisting, whose generator was defined as $T$ above. 
The charges for the fermions fields are $t_A = 1/2$, $t_B = -1/2$, 
$t_C = t_D =0$. The index theorem says that the difference between
the   number of zero modes on $\Sigma_g$ 
with positive and negative chirality is 
\eqn{numberfer}{
n^+ - n^- = { 1 \over 2 \pi} 
t \int_\Sigma F = {t \over 2 \pi } \int_\Sigma R = 2 t (g-1) ~, 
}
where we used that the gauge connection is the same as the spin connection.
 Since the ten-dimensional chiralities of the 
gauginos are positive, and we have taken them to have positive $SO(6)$
chirality, then we conclude that the chiralities in $1+1$ dimensions
are the same as those in $\Sigma$. So we conclude that the difference
between the number of left and right movers $n_R-n_L$ with given 
charges is also given by \numberfer .
Now it is useful to define $U(1)_L$ and $U(1)_R$ symmetries that
will become the $U(1)_L$ and $U(1)_R$ symmetries of the  CFT in the IR. 
In order to do this we note that the right-moving preserved supercharge
has charges as those of fermions in $A$, while the left-moving one
has charges as those in $B$. So we define $T_R = \half T_1 + \half T_2 +
T_3$
and $T_L =  -\half T_1 - \half T_2 + T_3$ with this definition we 
see that the supersymmetry parameter generating the right moving 
supersymmetries has charge $(q_R,q_L) = (1,0)$ and vice versa for 
the left moving one. These symmetries are anomalous and their anomalies
are given by the usual formulas $k_R = \sum_i (n^i_R - n_L^i) (q^i_R)^2$
where the sum runs over all fermions in the theory. 
Using \numberfer\ we get $k_L=k_R = N^2(g-1)$. Since the anomalies should
be the same at weak and strong coupling, and since  the (2,2) superconformal
algebra links the anomaly to the central charge we can compute the
central charge in the field theory to be 
\eqn{cetr}{
c_L =c_R = 3 N^2(g-1) ~.
} 

We can find the supergravity solution by the method we explained above.
We start by writing the supersymmetry variation equations, see the
appendix,
\eqn{eq511}{
g' = -\frac{1}{3}e^{f} [  (2 e^{-\varphi} +
e^{2 \varphi}) -  e^{-2g}   e^\varphi]}
\eqn{}{
f' = -\frac{e^f}{6} [2   (2 e^{- \varphi} +
e^{2\varphi}) + e^{ -2g}  
e^{\varphi}]}
\eqn{eq5111}{
\varphi' = \frac{2}{3}e^{f} [  (- e^{- \varphi} +
e^{2\varphi}) - \frac{1}{4} e^{-2g}  
e^{\varphi}]}

We could not solve these  equations  completely. However we could 
partially solve them by finding an analytic expression relating $g$ and 
$\varphi$
\eqn{relat}{
e^{2 g + \varphi} = e^{ 2g - 2 \varphi} + { g + 2 \varphi \over 2 } + C_1
}
When  $C_1 = 1/4 $ 
 the solution ends at the   fixed point where  $g'=\varphi' =0$. This 
corresponds to an $AdS_3 $ region where the Riemann 
surface has constant size. We can see from the equations
(\ref{eq511}) that this happens when $e^{2 g} =2^{-\frac{4}{3}} $ and $
e^{3 \varphi}
= 2 $. We also have then $e^{2 f} = { 1 \over 2^{4 \over 3} r^2}$. 
The solution 
interpolates between $AdS_5$ and $AdS_3 \times \Sigma$. 
 We can calculate the central charge from the
supergravity solution by calculating the effective three dimensional 
Newton's constant and then use the formula 
$ c = { 3 R_{ads} /( 2 G_N^{(3)}) }$.
In order to do this it is usefull first to calculate the five 
dimensional Newton's constant and then reduce from five
to three dimensions. We  are quotienting
$H_2$ so that we have a finite volume Riemann surface. 
Since the surface has constant curvature we can relate its volume
to the radius of curvature of $H_2$ and the genus. 
After we do this we get a central charge which agrees with \cetr .
See the appendix  for more details. 

Once we know the five dimensional metric and gauge fields it is easy 
to lift up the solutions to ten dimensions using the formulas in 
\cite{Cvetic:1999xp}. 
 The ten dimensional metric in the $AdS_3$ region, where $\varphi$ and $g$
are
constant,  reads
\ba
& &ds^2 = 2^{-{3\over 2}} \sqrt{\Delta} ds_3^2
+ { 2^{- {3 \over 2}}  \over  \sqrt{\Delta} } \left[ 
\Delta   { dx^2 + dy^2 \over y^2} + 2 \Delta d \theta^2 + 
\right. \nonumber \\ 
& & \left. + 4 \cos^2\theta (
d\psi^2 \!
+ \sin^2\psi\!(d\phi_1 + \frac{1}{2y}dx)^2 \! + \cos^2\psi\! (d\phi_2 +
\frac{1}{2y}dx)^2))\! +\!
  2  \sin^2\theta d\phi_3^2 \right] 
\label{tenmetric}
\ea
where we have defined;
\eqn{}{
ds_3^2 = { (dr^2 - dt^2 + dz^2) \over r^2 } 
}
\eqn{}{
\Delta= 1 + \sin^2\theta
}
the angles $\theta,\psi,\phi_i, i:1,2,3$ were introduced in order to
parametrize
the five sphere. The expression for
the self dual five form $F_5$ can also be obtained  from the formulas
in \cite{Cvetic:1999xp}. The metric  
 (\ref{tenmetric}) has  $ SO(2) \times SO(2) \times SU(2)$ isometries
as expected. The $SU(2)$ isometry  is not so obvious in the coordinates
in (\ref{tenmetric}) but can be made more manifest by writing
the original  $S^3$ parametrized by $\psi, \phi_1,\phi_2$ as the 
Hopf fibration on $S^2$. Then the gauge field is only appearing in the
fiber and the $SU(2)$ isometries are those of $S^2$.

It is interesting to note that there are other solutions like the one above,
where we allow a second scalar field $\varphi_2$.
 This second scalar field
is associated in the original four dimensional field theory to the 
operator ${\cal O}'_2 \sim |{\cal W}_1|^2 -|{\cal W}_2|^2$. 
In the IR region they describe  
marginal deformations of the theory with $\varphi_2 =0$. This  marginal
deformation breaks the $SU(2)$ isometry mentioned above to $SO(2)$. 
In the $AdS_3$ region the fields take values that can be parametrized
in term of $\varphi_2$ as 
(see the appendix  for details)
\eqn{solnew}{ e^{3 \varphi_1} = 2 \cosh \varphi_2
 \ \ \ \ \ \ e^{2g} = \frac{e^{2\varphi_1}}{4} \ \ \ \  \ 
f= - \log[ e^{2\varphi_1} r] 
}
It is easy to see that 
 the supergravity central charge  is the same as above \cetr .

There are more twistings of the D3 brane theory 
that we could consider. 
In the appendix  we discuss more general cases. In some of those
cases the spins of fields after twisting are not integer of half integer,
so it is not clear to us that the theories really make sense. 

Some twisted solutions of  five dimensional gauged
supergravities
were considered in \cite{Chamseddine:1999xk,Klemm:2000nj}.
 Those solutions had the restriction that
the scalar fields had to be constant in the flow from five to three 
dimensions. This restriction seems hard to obey if we insist in
not having fields with fractional spins, at least for the case of a 
D3 brane.

\section{Twists of the M5 brane theory}

In this section we consider $M5$ branes wrapped on two dimensional 
Riemann surfaces $\Sigma_g$ times $R^4$. 
We take the  limit $l_p \to 0$ keeping the size of the Riemann surface
fixed so that we obtain the $(0,2)$ six-dimensional superconformal
theory on $R^4 \times \Sigma_g$. We will consider two possible
ways of embedding the spin conection in the $SO(5)$ R-symmetry group
of the $(0,2)$ SCFT. In the first we preserve two supersymmetries
and in the second we preserve one supersymmetry (in $d=4$ notation). 
It is convenient to use a $U(1) \times U(1)$ truncation of 
the $SO(5)$ gauged supergravity theory. If we denote by 1,...,5 the 
directions orthogonal to the 5-brane, then these two $U(1)$'s rotate
the directions $12$ and $34$ respectively. This truncation 
was studied in \cite{Cvetic:1999xp} and the supersymmetry variation was studied
in \cite{Liu:1999ai}. 
In both cases we will make the following ansatz for the metric
\eqn{metricse}{
ds^2 = e^{2f(r)}(dr^2 + du^2 + dv^2 + dz^2 -dt^2) +
\frac{e^{2g(r)}}{y^2}(dx^2 +
dy^2)}
The boundary condition is as in the previous case that $g \sim f \sim 
- log(r) $, for $ r \to 0$. 

Throughout this section we work in units where the radius of 
$AdS_7$ is one. We can restore the dependence on this parameter
by multiplying the seven or ten dimensional metric by the radius
of $AdS_7$ which is given 
by $R^2_{AdS_7} = 4  (\pi N)^{2/3} l_p^3.~ $\footnote{
In conventions were the eleven dimensional Newton's constant
is $G_N^{11} = 16 \pi^7 l_p^9$  \cite{Maldacena:1998re}.}

\subsection{${\cal N}=2$ case}

In this section we consider configurations preserving 
${ \cal N} =2$ supersymmetry in four dimensions. 
The field content and
the 
Lagrangian of the $d=7$ gauged supergravity, together with the relevant
supersymmetry transformations are
described in detail in  \cite{Liu:1999ai} and in  the appendix.

In order to get an idea of what the field theory looks like let us
first consider a single M5 brane. 
In this case, in analogy with the D3 brane case we expect that 
there is an operator ${\cal O}_4 =  |{\cal Z}|^2 - { 2/3} 
({\cal W}_3^2 +{\cal W}_4^2+{\cal W}_5^2)$
where ${\cal Z} = {\cal W}_1 + i {\cal W}_2$ 
is the scalar that we are twisting and ${\cal W}_i$ are
the five scalar fields describing fluctuation of the M5 brane in the 
five orthogonal dimensions. 
The presence of this operator can be  seen by  expanding directly the 
Nambu action for the brane.  We can calculate its coefficient as
we did for the D3 brane case around \cuad , \cuadnew .
In the case of multiple coincident branes we also  have
an operator of dimension 4, with the same $SO(5)$ transformation
properties which is turned on. 
 So we expect  that a supergravity field, dual to this
operator,  will have a 
non-trivial boundary condition at infinity. 

In order to get an idea of what the field theory looks  like, let us
consider first the case of a single M5 brane. 
If it is wrapped on $S^2$ then we only get four massless modes, the
three scalar fields in the directions that are untwisted and 
the component of the $B$ field on  $S^2$. These form a single 
four-dimensional hypermultiplet.
This theory therefore  has only  a ``higgs'' branch. 
In the case that the M5 wrapps a genus $g>1$ Riemann surface 
then we get the same hypermultiplet that we were getting above plus
$g$ vector multiplets. These vector multiplets correspond to modes
of the $B$ field with one index along a non-contractible cycle in 
the Riemann surface and one index in four dimensions. 
So we get a $U(1)^g$ theory with one neutral hypermultiplet.
The scalars in the vector multiplet correspond to modes of the twisted
scalar field ${\cal Z}$ and they represent deformations of the 
Riemann surface in the normal directions that are twisted. 

It is less clear what the resulting four dimensional gauge theory 
is when we have $N$ coincident branes, since we do not have
an explicit Lagrangian we could use in six dimensions to derive
the four dimensional gauge theory. It seems possible to give 
a DLCQ definition of the theory in the spirit of 
\cite{Aharony:1998th,Aharony:1997an,Kapustin:1998pb}. 
Of course, it
would be very nice to give a direct definition of the theory.
As a step in that direction
we  compactify one of the worldvolume directions on a circle 
so that we get a D4 brane wrapped on $\Sigma$. Then the low energy
theory is a  three dimensional
$U(N)$ gauge theory with 8 supercharges.
 In the case of $S^2$ get get a pure gauge theory with only a Coulomb
branch 
while in the $g>1$ case we also get  $g$ adjoint hypers. 
Notice that the four dimensional theory we want to find should be such
that when we reduce it on a circle the Higgs branch of the 4d theory should
become the Coulomb branch of the 3d theory  and the Higgs branch of the
3d theory results from the Coulomb branch of the 4d theory as in 
\cite{Seiberg:1996nz}. After dimensional reduction to three dimensions
Higgs and Coulomb branches would be exchanged 
\cite{Intriligator:1996ex}.
In the case of $S^2$ we could say a bit more about this four dimensional
low energy theory. When we go down to three dimensions
we get a $U(N)$ field theory. The Coulomb branch of this theory
is given by the moduli space of $N$ monopoles in $SU(2)$ 
\cite{Seiberg:1996nz,Hanany:1997ie}. Three dimensional mirror symmetry
exchanges this Coulomb branch with the Higgs branch of the mirror. 
So the four dimensional theory is a theory whose Higgs branch is the 
same as the moduli space of $N$ monopoles of $SU(2)$. This space is smooth
and has no singularities. So the four dimensional theory 
is just this sigma model with only hypermultiplets \footnote{We thank
M. Douglas for suggesting this possibility.}. 
It is harder to say what the four dimensional field theory is in the 
case of $g>1$. Again we can say fairly easily what the Higgs branch 
of the four dimensional theory is. It is the same as the Coulomb branch
of the 3d  theory. It is less clear what the Coulomb branch of the 
four dimensional theory is. What we know is that upon reducing to
three dimensions and  doubling the
variables, as in \cite{Seiberg:1996nz} we should find the Higgs branch
of the D4 theory. The dimensionality of this Higgs branch is what
we analyzed for the D3 brane in section 3.1, it has dimension
$4 N^2(g-1)$. So we expect that the Coulomb branch of the 
four dimensional theory has dimension $ 2 N^2 (g-1)$. This is suggestive
of a very large gauge group of the form $U(N)^{(g-1)N}$.\footnote{ 
For a   related discussion see  \cite{Alishahiha:1999ds}.}
Below we will find, from a supergravity analysis, that the theory
has a conformal fixed point where the effective number of 
degrees of freedom goes as $N^3(g-1)$, which is in rough agreement with
the type of gauge group we expect.

Let us now turn to the supergravity solutions. 
Setting to zero the  supersymmetry variations we get
\eqn{gsev}{\eqalign{
g'=& -e^f [ e^{2 \lambda }
- \kappa \frac{e^{-2g}}{4} e^{3 \lambda } ]
+ \half \lambda'
\cr
\lambda '= & - { 2 \over 5} e^f [2 (e^{- 3 \lambda}
-  e^{2 \lambda} ) + \kappa \frac{1}{4}  e^{-2 g + 3 \lambda }]
\cr
f'= & -  e^{f + 2 \lambda }
+ \half \lambda'
}}
%
 where $\kappa = 1$ for $H^2$ and $\kappa = -1$ for $S^2$. 
The general solution of the equations is 
\eqn{solmfi}{\eqalign{
e^{5 \lambda} = & { e^{ 2 \rho} + \kappa  \half + C_1 e^{- 2 \rho} \over 
e^{2 \rho } + \kappa  {1 \over 4} }
\cr
e^{2 g} =& e^\lambda (e^{2 \rho} + \kappa  { 1\over 4} ) 
\cr
e^{2 f} = & C_2 e^{2 \rho} e^\lambda
\cr
e^{2 f} \left({dr \over d\rho}\right)^2 =&  g_{\rho\rho} = e^{-4 \lambda}
}}
We have chosen a new variable $\rho$ as implicit in the last
equation. $\rho \to + \infty$ corresponds to the boundary of
$AdS_7$.
$C_2$ is a trivial integration constant that can be absorbed by 
rescaling the four dimensional coordinates. It is  related
to the size of the surface where we are wrapping the six dimensional theory.
In the case $C_1=0$ $\kappa =1$  we find that the solution 
interpolates
between $AdS_7$ and $AdS_5 \times \Sigma_{g>1}$.  
This is telling us that the IR dynamics of the $(0,2)$ six dimensional
theory on $\Sigma_{g>1} \times R^4$ is given by a four dimensional 
superconformal field theory. The solution in the IR has the 
fixed point values
\eqn{fixedval}{
 e^{5 \lambda} = 2 ~,~~~~~~~ 
e^{2 g - \lambda} = {1 \over 4}~,~~~~~
e^{f + 2  \lambda} = { 1 \over r} ~.
}
The sign of the expectation value of the operator ${\cal O}_4$ is again given 
by the sign of $C_1$. A positive value of $C_1$ therefore corresponds 
to a positive expectation value for the twisted scalars and therefore
to the Coulomb  branch. A negative value of $C_1$ corresponds 
to an expectation value for the scalars that are untwisted and therefore
to the Higgs branch. 
However for the case of $S^2$ we do not expect a Coulomb branch since
the twisted scalars cannot have an expectation value. 
Fortunately either Gubser's criterion or our criterion in section 
\ref{criterion} 
rules out  this
case  but allows all the other cases. So in the case of $H^2$ we
can have both signs on $C_1$ and in the case of $S^2$ we find
that  $C_1 \leq {1 \over 16}$. The fact that we have $1/16$ instead
of zero is due to the fact that it is difficult to disentangle the 
expectation value from the insertion of the operator in supergravity 
so that our field theory expectation are easy to check only  for
large values of $|C_1|$, presumably a more correct analysis valid also
for small values of $C_1$ would still agree with field theory expectations.

%
Using the formulas in \cite{Cvetic:1999xp} we can uplift
the solution to eleven dimensions. Here we give only the 
form of the solution in its IR fixed point, the $AdS_5$ region.
\eqn{elev}{\eqalign{ ds_{11}^2 &= \half \Delta^{1/3} ds^2_{AdS_5} + 
{ \Delta^{ - {2 \over 3}} \over 4 } \left[ { \Delta }
 {( dx^2 + dy^2) \over y^2} + \right.
\cr
 & \left. + \Delta  d\theta^2 +  \cos^2 \theta (
d\psi + \sin^2 \psi d\phi^2_1 ) +2  \sin^2 \theta ( d\phi_2 + { dx \over y} )^2
\right]
\cr
\Delta &=  ( 1 + \cos^2 \theta) }
}

Here, the angles $\theta, \psi, \phi_1,\phi_2$ parametrize the four sphere
(before the twisting). $ds^2_{AdS_5}$ denotes a unit radius Anti-de-Sitter
metric. Note that \elev\ has $SU(2)\times U(1)$ isometries as required by
${\cal N} =2$ superconformal invariance. 
An explicit expression for the four form
gauge field can be written following section 4 of \cite{Cvetic:1999xp}.

This is giving non-singular  $AdS_5$ warped  compactifications of M theory. 
These compactifications are completely smooth. Singular compactifications
to $AdS_5$ were given in \cite{Alishahiha:1999ds,Cvetic:2000cj}, see 
however 
\cite{Fayyazuddin:1999zu,Fayyazuddin:2000em}. 

It is also possible to find the supergravity duals of the D4 brane
theories wrapped on $\Sigma$  that we discussed above by compactifying the 
M5 brane solution along a longitudinal direction and then using
the standard reduction from eleven dimensional supergravity to 
type IIA supergravity. We write this solution explicitly in the appendix. 
In the case of $S^2$ this solution is related to pure 3d SYM with 8 
supercharges. This is the same system that was explored in 
\cite{Johnson:1999qt}. In \cite{Johnson:1999qt} the starting point was
a D6 brane wrapping K3. At low energies this reduces to pure SYM in 3d. 
A puzzling aspect of the solutions in \cite{Johnson:1999qt} was that
the K3 remained in the IR geometry, at least at first sight. If we take
a limit of the solution in \cite{Johnson:1999qt} where the volume of 
K3 goes to zero, then we can T-dualize and obtain a large K3 and the
D6 brane becomes a D4 brane wrapping $S^2$ which  is what we have here.
In the IR region of our solution with $C_1 = 1/16$ we find that the
volume of $S^2$ shrinks to zero much faster than $g_{00}$ 
so we expect that KK excitations in this $S^2$ decouple from the low energy
theory. More precisely, dual of large $N$ pure SYM in 3d would be the
region close to the singularity. It is possible that there is a better
gravity description for the IR region 
than the one we explored here, or 
it could be that the theory only admits a string theory dual but not
a weakly coupled gravity dual. The excision of the singularity in 
\cite{Johnson:1999qt}  amounts to our criterion that the
singularity is admissible by the criterion in section \ref{criterion}. In fact
it was argued in \cite{Johnson:1999qt} that it was necessary to 
move the branes into the Coulomb branch of the 3d theory 
to remove the singularities.


\subsection{${\cal N}=1$ case}

In this section we consider the case where we embed the spin connection
in both of the $U(1)$ factors described above. 
When we wrap an M5 brane on a Riemann surface in a CY space we typically
get normal bundles which are topologically equivalent to this one. 
The resulting four dimensional field theory has 4 supercharges. 
It seems hard to say explicitly what this four dimensional theory is. 
Again if we compactify one circle and we go to three dimensions
we can say more. In that case we have a D4 brane wrapped on 
the Riemann surface. In the $S^2$ case we find just pure $U(N)$ gauge
theory with 4 supercharges.

Let us turn to the supergravity analysis.
The equations that result from setting to zero the supersymmetry
variations are (see the  appendix),
\eqn{fseve}{\eqalign{
f'=& -  e^{4\phi +f}  + \phi' 
\cr
g'=& -  e^{f}(  e^{4\phi} - \kappa {1 \over  4} e^{\phi -2g})  +\phi'
\cr
 \phi'=& -\frac{1}{5}e^f (4 (e^{-\phi} - e^{4\phi}) +  \kappa {1 \over  4}
e^{\phi - 2 g})
}}
where $\kappa = 1$ for $H^2$ and $\kappa =-1$ for $S^2$.
In the case of $H^2$, this 
set of equations  have solutions for constant values of
 $\phi, g$ 
given by
%
\eqn{solusev}{
e^{ 5 \phi }= \frac{4}{3}~,~~~~~\;\;\; e^{-2g - 3 \phi } = 4~,~~~~~
\;\;\; e^{f(r)} = e^{-4\phi} { 1 \over r} ~.
}
The corresponding eleven dimensional metric is smooth and 
can be obtained from 
\cite{Cvetic:1999xp}. Here we give only the $AdS_5$ region of the solution
which can be found using the IR values \solusev\
\eqn{11dmetricn=1}{\eqalign{
ds_{11}^2 = &\Delta^{\frac{1}{3}} ds_7^2 +
\frac{\Delta^{-\frac{2}{3}}}{4} [e^{-4 \phi} d\mu_0^2 + e^{\phi}
(d\mu_1^2 + d\mu_2^2 + \mu_1^2 ( d\phi_1 +\frac{1}{2y} dx)^2  
+ \mu_2^2 (d\phi_2+ \frac{1}{2y} dx )^2)]
\cr
\Delta=& e^{4 \phi} \mu_0^2 + e^{ - \phi} (\mu_1^2 + \mu_2^2)  
}}
where $\mu_i$ parameterize an $S^2$, $\mu_0^2 + \mu_1^2 + \mu_2^2 =1$
and the seven dimensional metric is as in  \metricse\ 
with the values given in 
\solusev .


It would be nice to determine whether these $AdS_5$ compactifications
of M-theory are stable under quantum corrections. In other words, it
would be interesting to determine whether these field theories are exactly
conformal of if they are non-conformal when we take into account 
$1/N$ corrections, as the ones in \cite{Kachru:1998ys}.

\section{ A criterion for allowed singularities}
\label{criterion}

In this section  we discuss  a proposal for a 
criterion that would tell us if a supergravity singularity in the
IR region of a geometry describing a field theory  is 
allowed or not. 
The strong  form of the final criterion is 

{\it The $g_{00}$ component of the metric should not increase
as we approach the singularity}

we will also discuss a weak form of the criterion 

{\it The $g_{00}$ component of the 
metric should be bounded above.}

In what follows we explain this criterion and give a heuristic
motivation for it. 
When we are trying to find supergravity solutions that are dual 
to field theories we typically encounter singular solutions.
These singularities do not necessarily mean that the solutions
 are wrong, they might be 
telling us that the supergravity description
is failing and that we should go to a dual description. 
This is the case for $D-p$-branes for  $p\not =3$ \cite{Itzhaki:1998dd}.
It is clear however that not all singularities are allowed. 
For example a solution like negative mass Schwarzschild should not
be allowed since it would imply that the energy is not bounded
below. We would like to propose a necessary criterion for weeding
out unphysical singularities. 
Our criterion  applies for the IR regions of supergravity backgrounds
which are dual to some field theory. 
If we want to interpret some region as being dual to the IR of some
field theory we expect that $g_{00}$ should decrease so
that fixed proper energy excitations correspond to lower and lower
energy excitations from the point of view of coordinate time,
which is the same as field theory time. 
So our criterion will be that $g_{00}$ should not increase as we 
approach the singularity. In particular it should not go to infinity.
In many cases we can approach the singularity in various directions
in the internal manifold. We require that $g_{00}$ does not increase 
as we approach the singularity 
along any direction    in the internal manifold. 
Note that it makes sense to talk about $g_{00}$ since we are talking
about a field theory with a time translation isometry, which is generated by 
a Killing vector of the dual geometry,
 so we are choosing coordinates so that this vector
is ${\partial \over \partial t }$.

This criterion certainly forbids 
negative mass Schwarzschild singularities
 where
$g_{00} \to \+ \infty$. 
There are some singularities where $g_{00}$ stays constant, like
 orbifold singularities. We should certainly allow those. Or 
course the full string theory will then tell us whether  it is really allowed
or not.  
There are some singularities where $g_{00}$ increases as one approaches
the singularity but stays bounded. A D8 brane has a singularity of this
form and it looks like it should be allowed in the full string theory 
but it seems that it should not be allowed in a region that we want
to interpret as the IR of some field theory.
 But in order to be sure
we are not  over-restrictive we could allow these and we get the weak form
criterion stated in the beginning of this section. If we rule these
out we get the first. 
A diverging $g_{00}$ implies
 that the singularity is repulsive, massive particles
are repelled from the singularity. In these singularities a 
finite proper energy excitation will have very high energy from 
the point of view of the field theory. It seems to violate the 
UV/IR correspondence.  It is also hard to see how these singularities
can form from gravitational collapse since they are repulsive. 
For a concrete example where an example of a  repulsive  singularity is 
discussed see \cite{Johnson:1999qt}. 
Note that  the $c$ theorem in 
\cite{Freedman:1999gp} does not imply that  $g_{00}$ always decreases. 
In \cite{Freedman:1999gp} it was
proven that the metric of gauged supergravity decreases, but
the full ten or eleven dimensional $g_{00}$ factor includes a warp
factor and this warp factor  could 
increase even though the metric appearing in the gauged supergravity
action is decreasing. 
The physical redshift factor is given by the ten or eleven dimensional
metric, as long as the supergravity solution is valid. In string theory
 one should use the Einstein metric for this analysis.

In \cite{Gubser:2000nd} it was proposed  that admissible singularities
are those that can be generalized to finite temperature. 
The criterion proposed here is easy to check and
it  can be applied directly to the 
ten or eleven dimensional metric and does not need to 
use the gauged supergravity form of the potential. 
In the cases analyzed here it gives the same results as 
Gubser's 
 criterion \cite{Gubser:2000nd} we suspect that the boundedness of the 
gauged supergravity potential will ensure that $g_{00}$ is bounded
but we did not prove it. 

Note that  this criterion is valid for regions that
we want to interpret as the IR of a gauge theory. It is certainly 
violated when we approach the boundary of $AdS$ or  the 
boundaries of the geometries describing some D-$p$-branes 
\cite{Itzhaki:1998dd}. In fact it is necessary
to put boundary conditions for fields in the boundary of $AdS$. These
boundary conditions are interpreted as defining the details 
(operator insertions) of the field theory dual to the given background.

\section{A no go theorem}
\label{nogo}

In this section, which is to great extent disconnected with the
rest of the paper, we present an argument  saying that there
are no non-singular wrapped compactifications in a large class 
of supergravity theories. Our main assumption will be that the
potential for scalar fields is non-positive. The massive type IIA case
is treated separately.  We do not need to use the equations of motion
of the matter fields. 
Then we will show that there is no non-singular warped compactification
to $R^d$ or  de-Sitter space $dS^d$, $d\geq 2$ with finite $d$ dimensional
Newton's constant. We will do this for general $d$, but the reader
interested in real world applications might want to  take $d=4$. 
The no go theorem remains true even in the case that we allow
singularities such as those allowed by the strong form of the criterion
in section \ref{criterion}. These are singularities that we {\it might} be able
to interpret as arising from the IR dynamics of some field theory. 
The argument is quite general and only relies on the equation of
motion for the warp factor and does not rely on supersymmetry. 
There are ways to evade this argument which involve including higher
derivative corrections to the supergravity equations, or starting
from a theory that already has a positive cosmological constant.
There are  other no  go theorems for supersymmetric RS 
models using five dimensional gauged supergravity 
\cite{Wijnholt:1999vk,Kallosh:2000tj,Gibbons:2000hg}. These are
complementary
to our arguments, since the five dimensional potential could be positive.
 If the five dimensional gauged supergravity arises from
a large volume compactification then we could apply our results but 
the arguments in  \cite{Wijnholt:1999vk,Kallosh:2000tj,Gibbons:2000hg}
 also cover 5d theories  which cannot be interpreted as large volume 
compactifications.

 We consider a $D$ dimensional gravity theory, with $D>2$,  compactified 
down to $d$ dimensions. We denote by $M,N,L,..$ the $D$ dimensional
indices. We denote by $\nu,\nu,\rho,...$ the $d$ dimensional indices and
by $m,n,l,$ the $D-d$ dimensional indices. 
We will assume that the $D$ dimensional gravity theory satisfies the
following conditions. 
\begin{itemize}
\item
The gravity action does not contain higher curvature corrections.
\item 
The potential is non-positive, $V\leq 0$.  This condition in not obeyed in
massive IIA supergravity which has a positive cosmological constant so we
treat
that case separately in \ref{massivetwoa}.
 $V$  
 could be just a  negative cosmological constant or it can
depend on the scalars but it cannot be positive (at least in the 
range  of values of scalar fields that is  explored in the 
solution under consideration). 
\item
The theory contains  massless fields with positive kinetic terms. These 
massless fields have field strengths which are $n$ forms, $F_{i_1,..,i_n}$.
For $n=1$ we have scalar fields, $n=2$ Maxwell fields (these could
be non-abelian, as long as the metric on the group is positive definite
so that the kinetic terms are positive), etc. We consider $n < D$, if
$n=D$ it would give a contribution similar to a potential and we go back
to the previous assumption.
\item
The $d$ dimensional effective Newton's constant is finite.
\end{itemize}

We start by writing out Einstein's equations in $D$ dimensions
\eqn{einst}{
R_{MN} = T_{MN} - { 1 \over D -2}g_{MN} T^{L}_{~L}
}
Notice that in \einst\ we neglected higher derivative corrections. 
We write the metric as 
\eqn{ansmet}{
ds^2_D = \Omega^2(y) \left( dx^2_{d} + \hat g_{mn} dy^ndy^m \right)
}
where $dx^2_d = \eta_{\mu\nu}dx^\mu dx^\nu$ where $\eta$ is the 
metric of the $d$ dimensional space which is  either 
  Minkowski or de-Sitter space. 
Now we calculate the $R_{\mu\nu}$ components of the $D$ dimensional
metric and we find that Einstein's equations imply
\eqn{einst}{
R_{\mu\nu} = R_{\mu\nu}{(\eta)} - \eta_{\mu\nu} 
\left(  \hat \nabla^2 \log \Omega + (D-2) 
(\hat \nabla \log \Omega )^2  \right)  = 
T_{\mu\nu} - { 1 \over D-2} \Omega^2 \eta_{\mu\nu} T^L_{~L}
}
where the hat denotes covariant derivatives and contraction of indices 
with respect  to the metric
$\hat g$.
Taking the trace over $\eta$ on both sides we find
\eqn{trace}{
\hat \nabla^2 \log \Omega + (D-2) ( \hat \nabla \log \Omega )^2 
= {1 \over(D-2)  \Omega^{D-2}} \nabla^2 \Omega^{D-2} = 
 R(\eta) + \Omega^2 (  - T^\mu_{~\mu} + { d \over D-2}  T^{L}_{~ L} )
}
where in the term involving the stress tensor on  the right hand side
 we contract the indices with the $D$ dimensional
metric and $R(\eta)$ is the curvature of the $d$  dimensional metric $\eta$.
We will now proceed to prove that the term in the right hand side involving
the stress tensor is non-negative. 

\subsection{ $\tilde T \geq 0$ }

The stress tensor will be the sum of the contributions to the stress 
tensor of the various massless fields.
 We will consider each contribution
individually since they are all adding up to the total stress tensor.
Let us define 
\eqn{tilten}{
\tilde T \equiv   - T^\mu_{~\mu} + { d \over D-2}  T^{L}_{~ L}
}
We want to show that all contributions to $\tilde T $ are non-negative. 
Let us first consider the potential term. 
We will not keep track of irrelevant positive numerical constants. 
The stress tensor is 
\eqn{pot}{
T_{MN} \sim - V g_{MN} ~, ~~~~~~~~ \tilde T \sim
 -V { 2 d \over D-2}  \geq
0
}
if  $V<0$ as assumed. 
Now let us consider the $n$ form field strengths. 
Their stress tensors are 
\eqn{tens}{ \eqalign{
T_{MN} = & F_{M L_1 ..L_{n-1} } F_{N}^{~L_1 ..L_{n-1}} - { 1 \over 2 n}g_{MN} 
F^2 \cr
\tilde T = & - F_{\mu L_1 ..L_{n-1}} F^{\mu L_1 ..L_{n-1}} + { d \over D-2}
( 1 - {1 \over n}) F^2 
} }
In principle we could have functions of scalar fields multiplying these
expressions, as we have in some supergravity theories, 
 and we could also have many types of $n$ form fields. 
We will  not indicate these explicitly but it is obvious how to extend
the following arguments to those cases. 
The space time indices of non-vanishing components of $F$ could be
completely along the internal dimensions or, if $n\geq d$,  they could have
$d$ out of $n$ indices along the $d$ dimensions and the rest along
the internal dimensions.  Other possibilities do  not preserve the
isometries of $R^d$ or $dS^d$. In constructing  $\tilde T$ these
two types of components will make separate contributions. We will therefore
consider them independently and show that each of them is positive. 
So let us  first consider the part of $F$ with all indices internal.
Then we have that $F^2 \geq 0$ and we see from \tens\ that we have a 
positive contribution.  For all $n>1$ forms this contribution is strictly 
positive if we have a non-vanishing field strength, but for $n=1$ the 
contribution is zero even if we have a non-vanishing field strength.
Now we consider the part of the field strength with components along
the $d$-dimensional space. 
The difference between the term that contains a trace over the $\mu$ index
and the others is that we are choosing a particular order of contractions
of the indices comparing the two we find that 
$$ F_{\mu L_1 ..L_{n-1}} F^{\mu L_1 ..L_{n-1}} = {d \over n} F^2 .$$
Then we find that 
\eqn{tilddim}{
\tilde T 
=  - F^2  { d ( D-1    - n   ) \over n( D-2)}  \geq 0
}
Where we used that $F^2 <0$ since we are considering temporal components 
of $F$. We have also used that we are considering $n \leq   D-1 $. 

\subsection{ Condition on the warp factor}

Multiplying   \trace\  by a power of $\Omega$  and using that $\tilde T
\geq 0$ we conclude that
\eqn{warp}{
\hat \Omega^{(D-2)} \nabla^2 \Omega^{(D-2)} \geq 0
}
with equality holding only if the right hand side of \trace\ is zero
so that  the $d$ dimensional space is Minkowski space.
Remember that the $d$ dimensional Newton constant is given by 
\eqn{ddimnew}{
{ 1 \over G_N^d } \sim \int d^d y \sqrt{ \hat g} \Omega^{(D-2)}
}
We are assuming that this Newton constant is finite. 

Let us first assume that $\Omega$  is bounded below  and above 
in the internal manifold.
In that case the internal manifold should be compact. 
Integrating  \warp\ over the compact internal space  by parts we 
conclude that 
$
\int d^(D-d) y \sqrt{\hat g}  ( \hat \nabla \Omega^{(D-2)} )^2 \leq 0 $ 
which is possible only if $\Omega$ is constant. 
In that case we conclude that the right hand side of \trace\ is zero,
so that we cannot have a deSitter space and the only $n$ forms that 
we can be  turned on are the $n=1, D-1$ forms. 

As discussed in section \ref{criterion}  we expect that singularities where 
$\Omega$ diverges should not be allowed. So we conclude that 
$\Omega$ is bounded above.
Now suppose that we have regions where $\Omega \to 0$ or we 
 allow singularities obeying the strong form
of the criterion in section \ref{criterion},
 which says that $g_{00}$ should not
increase as we approach the singularity.
In this case we can define a region ${\cal R}$ which leaves out the
singularities  and  such that
$\Omega > \epsilon$ in ${\cal R}$  for a suitably small $\epsilon$.
By our assumptions about the singularities it is clear that we can
choose ${\cal R}$ so that 
 $\nabla \Omega$  is either zero or pointing inwards
at the boundary of ${ \cal R}$. 

Now we can integrate \warp\ by parts in region ${\cal R}$,
and we get
\eqn{poswar}{
 \int_{{\cal R}_\epsilon} 
\left(\hat \nabla \Omega^{(D-2)} \right)^2 \leq - \int_{\partial 
{\cal R}_\epsilon}
( \vec n \nabla \Omega^{(D-2)} ) \Omega^{(D-2)} \leq 0
}
where we used the assumption that $\Omega$ was non-increasing at the boundary.
Again we conclude that the warp factor has to be constant and 
that a R-S or  deSitter compactification of this type is  not allowed.
\footnote{
Note that the solutions in \cite{Cvetic:1999fe} do 
not obey the condition
that
the higher dimensional $g_{00}$ goes to zero at the singularity.
}.

In summary we have proven that given our assumptions there are 
no compactifications to deSitter space or 
Randall-Sundrum compactifications where the only possible singularities
are such that the warp factor 
 $\Omega$ going to zero at the singularity. As we explained in 
section \ref{criterion},
 these are the singularities which {\it might}  have a field
theory interpretation.

The fact that de-Sitter is not allowed {\it does not}  imply that 
there are no expanding universe solutions in large volume compactifications,
it only 
says that there are no homogeneous $SO(1,d+1)$ invariant 
  de-Sitter solutions. So we could
have solutions where some scalar fields are time dependent.\footnote{
The same comment applies to the statement 
in \cite{Arkani-Hamed:2000eg}
 that their compactifications have no de-Sitter or 
Anti-de-Sitter solutions. On general grounds we  expect to find 
 also expanding or contracting universes with time dependent scalar
fields  containing
singularities similar  to the ones considered in \cite{Arkani-Hamed:2000eg}.}

The most natural question is whether there are ways to evade this 
argument. For that we note that once we include higher derivative
corrections to the gravity action, like the ones  present in 
string theory or M-theory, then the positivity argument does not
hold and there can be warped compactifications 
\cite{Strominger:1986uh,Witten:1996mz,Dasgupta:1999ss,%
Chan:2000ms,Behrndt:2000zh,Mayr:2000zd}. 
For example, in heterotic compactifications or Horava Witten
 compactifications \cite{Horava:1996qa} 
we can have warp factors but we need crucially the higher derivative
terms which modify the Bianchi identity for the three or four form
field strengths respectively.
 The same can be said about type I examples
where we have orientifolds. These higher derivative terms are 
crucial to get important physical aspects of string compactifications.
All we are saying is that these stringy corrections to the
gravity equations  are also crucial if one wants a 
deSitter of Randall Sundrum compactification \cite{Randall:1999vf}. 
Of course, it is  very interesting to study these 
solutions more precisely.  
It is possible to evade the arguments in this section by 
by allowing 
potentials for scalar fields that are positive, as explicitly 
demonstrated in \cite{Cvetic:1996vr,DeWolfe:1999cp}.

\subsection{ No Minkowski 
 or de Sitter compactifications of massive IIA sugra}
\label{massivetwoa}

In this section we show that there are no compactification of 
massive IIA on smooth manifolds without boundaries 
down 
to $R^d$ or $dS^d$. 
The equations of  motion for the metric and the dilaton in 
 massive IIA supergravity are given by \cite{Romans:1986tz}
\eqn{metrie}{\eqalign{
R_{MN} =& {m^2 \over 16}e^{-5 \phi} g_{MN} + 2 \partial_M \phi
\partial_N \phi + e^{2 \phi} ( G_M^{~PQ}G_{NPQ} -{ 1\over 12} g_{MN} 
G^2 ) + 
\cr
&~ +  2 m^2 e^{- 3 \phi} ( B_M^{~P} B_{NP} - {1 \over 16}g_{MN} B^2 ) 
+ { 1\over 3 } e^{ - \phi} ( F_M^{~PQR} F_{NPQR} - { 3 \over 32}g_{MN} F^2)
}}
\eqn{dile}{ 0 = 
\Box \phi + { 5 m^2 \over 8} e^{ - 5 \phi} + { 1 \over 48} e^{-\phi}
F^2 - {1 \over 6} e^{2 \phi} G^2 + { 3 m^2 \over 4} e^{ - 3 \phi} B^2
}
where we have defined the square of a tensor of $n$ indices as 
$H^2 = H^{M_1 ...M_n} H_{M_1...M_n}$. We will not need the precise 
definition of the tensors $B,~G$ and $F$ in \metrie \dile\ but 
we will  use that they are real and antisymmetric. We assume $d\geq 2$. 
We make an ansatz for the metric as in \ansmet . Following the steps
that lead to \trace\ we find 
\eqn{tracemass}{\eqalign{
& {1 \over(D-2)  \Omega^{D}} \nabla^2 \Omega^{D-2} = 
\Omega^{-2}
 R(\eta) +d \left[ -{m^2 \over 16}e^{-5 \phi}  + {1 \over 12}  e^{2 \phi}
G^2_e + { m^2 \over 8} e^{- 3 \phi} B^2_e 
+ { 1\over 32 } e^{ - \phi} F^2_e \right] - 
\cr
 & ~~~~~~~- d \left[ \theta(3-d) { 1 \over 4}
 e^{2 \phi}
G^2_l +  \theta(2-d) { 7 m^2 \over 8}  B^2_l + \theta(4-d){ 5 \over 96} 
 e^{ - \phi} F^2_l \right]
}}
where we have denoted by $H^2_e$ the square of a tensor with components
purely in the internal dimensions and by $H^2_l$ the square of the
tensor with components along the $d$ dimensional spacetime directions.
These can only appear if the rank of the tensor is bigger or equal
to $d$ that is the reason we have factors of $\theta(n-d)$ where 
$\theta(x) =1 $ for $x\geq 0$ and zero otherwise. 
Now notice that
\eqn{boxfi}{
\Box \phi = { 1 \over  \Omega^D} {\hat \nabla}_m \left( \Omega^{D-2}
 \hat g^{mn}
\partial_n \phi \right) 
}
We then see that if we multiply \tracemass\ by $10 \Omega^D$ and add
\boxfi\ times $d \Omega^D$ we get
\eqn{final}{\eqalign{
&  {10 }  \Omega^D \times \tracemass + d \Omega^D \times \dile =
 { 10 \over D-2} {\hat \nabla}^2 \Omega^{D-2} +
  d {\hat \nabla}_m \left( \Omega^{D-2}
 \hat g^{mn}
\partial_n \phi \right)= 
\cr
&~~~~= \Omega^{D-2} R(\eta) + 
+ d \Omega^D \left[  {2 \over 3}   e^{2 \phi}
G^2_e + { 2 m^2 } e^{- 3 \phi} B^2_e
+ { 1\over 3 } e^{ - \phi} F^2_e \right. - 
\cr
&~~~~~~~  \left. -\theta(3-d) { 8 \over 3}
 e^{2 \phi} G^2_l -  
\theta(2-d) { 8 m^2}  B^2_l - \theta(4-d){ 1 \over 2 } 
 e^{ - \phi} F^2_l \right]
}}
We see that the right hand side of \final\ is positive since $H_e^2 >0$ and
$H_l^2 <0$ for all $n$ forms.

If we  have a  compact internal manifold then
we can integrate \final over the manifold. We get zero from the left 
hand side since we have a total derivative. On the right  hand side we 
get a non-zero result unless $R(\eta) = 0$ and $ B=F=G=0$. Now that we know
this we can 
integrate just the equation for the dilaton \dile\ over the internal
manifold and
we  get a contradiction. So we find that there are no non-singular
compactifications (over a compact internal manifold) to either de Sitter
or Minkowski space. 
If we assume that we have regions where the warp factor could go to zero
but the dilaton stays constant, as we expect  in an AdS region. 
Then we also get 
a contradiction by following steps similar to those in the previous 
subsection and including boundary terms for the conformal factor. 
This excludes  compactifications of the RS type. 
In \cite{Romans:1986tz} there are several examples of compactifications
to Anti-de-Sitter manifolds. 

Obviously there are compactifications of massive IIA to Minkowski space 
on spacetimes with
boundaries  where the conformal factor is not decreasing as we 
approach the boundaries as explicitly demonstrated in
 \cite{Polchinski:1996df}. 
It would be nice to see if we can get compactifications to 
de Sitter space in cases where the compact manifold has boundaries.

\section*{Acknowledgements}

We would like to thank M. Bershadsky, 
M. Douglas, J. Gomis, K. Hori, C. Johnson,
 A. Kapustin, A. Strominger and
C. Vafa for discussions. 

The research of C.N. was supported by CONICET from Argentina and
the Antorchas foundation. 
The research of J.M.\ 
was supported in part by DOE grant DE-FGO2-91ER40654,
NSF grant PHY-9513835, the Sloan Foundation and the 
David and Lucile Packard Foundation.

\section{Appendix}

Here we discuss some details in obtaining the supergravity solutions.

\subsection{D=5}

We start with IIB supergravity in $AdS_5 \times S^5$. It is believed
that  we 
can obtain a consistent truncation to ${\cal N} =8$ supergravity.
${\cal N}=8$ supergravity involves an $SO(6)$ gauge field. 
We can further truncate this theory to supergravity theories
involving a smaller gauge group. The advantage of considering these
truncations is a simplification in the equations of motion.
The truncation that we used is a truncation to a supergravity
theory in five dimensions with three $U(1)$ gauge fields and two 
scalar fields described in \cite{Chamblin:1999tk},\cite{Cvetic:1999xp}.
 This is an ${\cal N} =2 $ supergravity theory in 
five dimensions with two vector multiplets. The third gauge field
is the graviphoton. The three $U(1)$ gauge fields are associated to 
rotations on the 12, 34 and 56 planes respectively, where 1-6 are
directions
orthogonal to the D3 brane.  

The Lagrangian for this theory is given by \cite{Cvetic:1999xp}
%
\cite{Chamseddine:1999xk}) is
\eqn{lagrfd}{
L= R - 
\frac{1}{2}(\partial_\mu \phi_1)^2
-\frac{1}{2}(\partial_\mu \phi_2)^2
+ 4 \sum_{i}^{3}e^{\alpha_i}  - \frac{1}{4}(\sum_{i}^{3}
e^{2 \alpha_i}{{F^{i}}_{\mu\nu}}^2) +
\frac{1}{4}\epsilon^{\mu\nu\alpha\beta\rho}{F^{1}}_{\mu\nu}
{F^{2}}_{\alpha\beta}{A^{3}}_\rho
}
where we have defined
\eqn{fieldsdef}{
\alpha_1 =
\frac{\phi_1}{\sqrt{6}} +\frac{\phi_2}{\sqrt{2}},\;\;
\alpha_2 =\frac{\phi_1}{\sqrt{6}} -\frac{\phi_2}{\sqrt{2}},~~
\alpha_3 =\frac{-2\phi_1}{\sqrt{6}},
}
We work in  units where  $R_{AdS_5} =1$ for
the usual  $AdS_5$ solution, related to ${\cal N}  =4$ SYM.
The corresponding supersymmetry
transformations are
\eqn{}{\eqalign{
\delta \lambda_i &= (\frac{3}{8}\partial_i X_I \Gamma^{\mu\nu} F_{\mu\nu}^I
- \frac{i}{2}g_{ij}\Gamma^\mu \partial_\mu \phi^j +\frac{3i}{2}V_I
\partial_i X^I)\epsilon
\cr
\delta\psi_\mu  & = (D_\mu +\frac{i}{8} X_I(\Gamma_\mu^{\nu\rho} - 4
\delta_\mu^\nu \Gamma^\rho )F_{\nu\rho}^I + \frac{1}{2}\Gamma_\mu X^I V_I
 - \frac{3i}{2} V_I A_\mu^I)\epsilon
}}
we have defined
\eqn{x}{X^{(1)}= e^{-\frac{\phi_1}{\sqrt{6}} -\frac{\phi_2}{\sqrt{2}}}
  \ ~~~~~~~~~~~
X^{(2)}= e^{-\frac{\phi_1}{\sqrt{6}} +\frac{\phi_2}{\sqrt{2}}}
  \ ~~~~~~~~~
X^{(3)}= e^{2\frac{\phi_1}{\sqrt{6}}}}
\eqn{}{
X_{(1)}=\frac{1}{3} e^{\frac{\phi_1}{\sqrt{6}}
+\frac{\phi_2}{\sqrt{2}}}\ ~~~~~~~~~~
X_{(2)}=\frac{1}{3} e^{\frac{\phi_1}{\sqrt{6}}
-\frac{\phi_2}{\sqrt{2}}}\ ~~~~~~~~~~
X_{(3)}= \frac{1}{3}e^{-2\frac{\phi_1}{\sqrt{6}}}}

\eqn{}{
g_{ij} = \frac{1}{2}\delta_{ij}
\;\;\; ~~~~~~~~~ V_I=\frac{1}{3}}
For the metric \ans , we have the spin connection
\eqn{}{
\omega_{\hat{x}}^{x y}=-\frac{1}{y} \;\;\;
\omega_{\hat{x}}^{xr}= \omega_{\hat{y}}^{yr}=\frac{e^{g-f}g'}{y}, \;\;\;
\omega_{\hat{z}}^{zr}= 
\omega_{\hat{t}}^{tr}= f',}
(here the hatted index is curved and the others are flat) and with the
choice of gauge fields
\eqn{}{A_{x}^{(1)}=\frac{a}{y}\;\;\;
A_{x}^{(2)}=\frac{b}{y}\;\;\; A_{x}^{(1)}= \frac{c}{y}}
we have the equations
\eqn{fone}{\partial_x \epsilon
+\frac{1}{2}\omega_{\hat{x}}^{xy}\Gamma_{xy}\epsilon
+\frac{1}{2}\omega_{\hat{x}}^{xr}\Gamma_{xr}\epsilon -\frac{i}{2}X_{I}
F_{xy}^{I} y e^{-g} \Gamma_y \epsilon +\frac{e^g}{2y}X^{I} V_I
\Gamma_x\epsilon -\frac{i}{2}(A_x^{(1)} + A_x^{(2)} +
A_x^{(3)})\epsilon=0} 
\eqn{ftwo}{\partial_z\epsilon +
\frac{1}{2}\omega_{\hat{z}}^{zr}\Gamma_{zr}\epsilon + \frac{i}{4}X_I
F_{xy}^I e^{-2g} y^2\Gamma_{zxy}\epsilon + \frac{1}{2}e^f X^I V_I
\gamma_z\epsilon=0}
\eqn{fthree}{\frac{3}{4} \partial_{\phi1}X_I F_{xy}^Ie^{-2g} y^2
\Gamma_{xy}\epsilon -\frac{i}{4} e^{-f} \phi_1' \Gamma_r \epsilon
+\frac{i}{2} \partial_{\phi1}(X^{(1)}+X^{(2)}+X^{(3)})\epsilon =0}
\eqn{ffour}{\frac{3}{4} \partial_{\phi2}X_I F_{xy}^Ie^{-2g} y^2
\Gamma_{xy}\epsilon -\frac{i}{4} e^{-f} \phi_2' \Gamma_r \epsilon
+\frac{i}{2} \partial_{\phi2}(X^{(1)}+X^{(2)}+X^{(3)})\epsilon =0}
and another equation describing the radial dependence of the spinor that
we do not write  here.
We impose the following condition on  the spinors
\eqn{condspinor5}{\Gamma_{xy}\epsilon= -i\beta\epsilon
~~~~~\Gamma_r\epsilon =\eta \epsilon }
where $\beta = \pm 1, ~\eta = \pm 1$. By doing simply parity transformations
we set $\beta = \eta =1$.
We obtain, from eq. \fone\
\eqn{}{1=  (a + b + c)}
\eqn{}{g' =-2 e^f [\frac{1}{6}(X^{(1)}+X^{(2)}+X^{(3)})
-\frac{1}{2}X_I a^I e^{-2g} ]}
Equations  \ftwo \fthree\ \ffour\ reduce to
\eqn{}{f'= -2 e^f [\frac{1}{6}(X^{(1)}+X^{(2)}+X^{(3)})
+\frac{1}{2}X_I a^I e^{-2g} ]}
\eqn{}{\phi_1'= 4e^f[\frac{1}{2}\partial_{\phi1}(X^{(1)}+X^{(2)}+X^{(3)}) 
-\frac{3}{4}\partial_{\phi1}X_I a^I e^{-2g}]
}
\eqn{ric}{\phi_2' 
= 4e^f[\frac{1}{2}\partial_{\phi2}(X^{(1)}+X^{(2)}+X^{(3)}) 
-\frac{3}{4}\partial_{\phi2}X_I a^I e^{-2g}]}
It is easy to see that, the choice 
\eqn{ph2}{\phi_2=0, a=b}
 solves  \ric .
Using the explicit expressions for the
scalar
fields $X^I$  and defining $\varphi
=\frac{\phi}{\sqrt{6}}$ we obtain
%
\eqn{eq51}{
g' = -\frac{1}{3}e^{f-g} [ e^g (2 e^{-\varphi} +
e^{2\varphi}) -  e^{-g} ( 2a e^{\varphi}
+ c e^{-2\varphi})]}
\eqn{}{
f' = -\frac{1}{6} [2  e^f  (2 e^{-\varphi} +
e^{2\varphi}) +  e^{f -2g} ( 2a
e^{\varphi}
+ c e^{-2\varphi})]}
\eqn{}{
\phi' = \frac{4}{\sqrt{6}}e^{f} [  (- e^{-\varphi} +
e^{2\varphi}) - \frac{1}{2} e^{-2g} ( a
e^{\varphi}
- c e^{-2\varphi})]}
\eqn{eqn54}{
2 a + c = 1}
In the case $ a=b=0;~c=1$
 we obtain eqs. \susycond , while
for $a=b=1/2 $ we obtain 
eqs. (\ref{eq511}) - (\ref{eq5111}) .

It can be seen that  solutions with constant
$g,\phi$ can be obtained
\eqn{solutionscte}{\eqalign{
a=& b ~,~~~~~~~~
c= 1 - 2 a ~,~~~~~~~ \;\;\; e^{\frac{\phi}{\sqrt{6}}} =( 4
-\frac{1}{a})^{\frac{1}{3}}
\cr 
 e^{-2 g} =&
\frac{ (4 - \frac{1}{a})^{\frac{1}{3}}}{a}
=\frac{e^{\varphi}}{a} 
\cr
e^f=& e^f_0 {1 \over r}~,~~~~~~~e^f_0 = 
\frac{2a (4 - \frac{1}{a})^{\frac{1}{3}}}{(6 a - 1)} 
}}

A simple analysis of the eqs. (\ref{eq51})-(\ref{eqn54}) near  $r= 0$,
shows that
\ba
& &g(r)= - Log(r) + \frac{7 (2 a + c)}{36}r^2 + ...\nonumber\\
& &f(r)= - Log(r) - \frac{ (2 a + c)}{18}r^2 + ...\nonumber\\
& &\phi(r)=  -\frac{ ( a - c)}{3}r^2 Log(r) + ...
\ea
The last equation shows that the operator dual to the field $\phi$ is
turned on. The field $\phi$ is dual to an operator of dimension $\Delta
=2$.
For this dimension the two solutions of the wave equation go as 
$\phi \sim r^2 $, $ \phi \sim r^2 \log(r)$. The second solution
is the non-normalizable mode associated to the insertion of an operator. 
We see that in both cases analyzed in this paper the operator $\phi$ is 
turned on. 
It is interesting that there is also a special solution where 
$a = b = c = 1/3$ where the operator $\phi$ is not turned on. The field
$\phi=0$ in the whole solution. This solution is of the form of the 
solutions analyzed in \cite{Klemm:2000nj}. It would be interesting to
see if this solution really makes sense since some fields acquire 
fractional spins with this twisting. 
In this case the metric has the form \cite{Klemm:2000nj}
\eqn{}{ds^2 = e^{-g} (3 e^{2g} -1)^{\frac{3}{2}}(-dt^2 + dz^2) +
\frac{9}{(3 - e^{-2g})^2} dg^2 +
\frac{e^{2g}}{y^2}(dx^2 + dy^2)}

We now  give  some details on the computation of the
central charge for the solutions of the form \solutionscte\
from the supergravity side. We have

\eqn{}{
c=\frac{3 R_{AdS3}}{2 G_N^{3}}}
and
\eqn{}{
{ 1 \over G_N^{3}} =\frac{ Vol_7}{G_N^{10}} = 
{ 2 N^2 \over \pi } Vol{H_2}
}
where we used that in the units used in our paper where $R_{AdS_5} = 1$
we have that $G_N^{{10} } = { \pi^4 \over 2 N^2}$ and that the
volume of a unit radius five-sphere is $\pi^3$. 

In the case of the solutions \solutionscte\ 
$R_{AdS3} = e^{f_0}$.
We can calculate the volume of a constant curvature Riemann surface
as follows
\eqn{}{
\int_{\Sigma_2}\sqrt{g}R= 8\pi(1-g)~; \;\;\; ~~~~~~
R=-2e^{-2g} ~~~~~ \to ~~~ \int_{\Sigma}\sqrt{g} = 4 \pi (g-1) e^{2 g}
}
\eqn{}{
c= 12 N^2(g-1) e^{2 g + f_0} =   \frac{24 a^2 N^2 (g-1)}{6 a-1}}
for the case $a=\half$ we reproduce \cetr\ .

As we pointed out in the paragraph before eq. \solnew\ there are other
solutions for the case in which we excite two gauge fields
$A_x^{(1)}=\frac{a}{y}, \;\; A_x^{(2)} =\frac{b}{y}$ and we keep both
scalar fields
$\phi_1,~\phi_2$ nonvanishing.

in that case the eqs. to solve read

\eqn{}{
g' = -\frac{1}{3}e^{f} [ ( e^{-\frac{\phi_1}{\sqrt{6}}
-\frac{\phi_2}{\sqrt{2}}} +
 e^{-\frac{\phi_1}{\sqrt{6}}
+\frac{\phi_2}{\sqrt{2}}} +  e^{2\frac{\phi_1}{\sqrt{6}}}) -  e^{-2g}
( a e^{\frac{\phi_1}{\sqrt{6}}
+\frac{\phi_2}{\sqrt{2}}} + b  e^{\frac{\phi_1}{\sqrt{6}}
-\frac{\phi_2}{\sqrt{2}}} + c  e^{-2\frac{\phi_1}{\sqrt{6}}})}
\eqn{}{
f' = -\frac{1}{6}e^{f} [ 2( e^{-\frac{\phi_1}{\sqrt{6}}
-\frac{\phi_2}{\sqrt{2}}} +
 e^{-\frac{\phi_1}{\sqrt{6}}
+\frac{\phi_2}{\sqrt{2}}} +  e^{2\frac{\phi_1}{\sqrt{6}}}) +  e^{-2g}
( a e^{\frac{\phi_1}{\sqrt{6}}
+\frac{\phi_2}{\sqrt{2}}} + b  e^{\frac{\phi_1}{\sqrt{6}}
-\frac{\phi_2}{\sqrt{2}}} + c  e^{-2\frac{\phi_1}{\sqrt{6}}})}
\eqn{}{
\phi_1' = \frac{1}{\sqrt{6}}e^{f} [2 (- e^{-\frac{\phi_1}{\sqrt{6}}
-\frac{\phi_2}{\sqrt{2}}}  - 
e^{-\frac{\phi_1}{\sqrt{6}}+\frac{\phi_2}{\sqrt{2}}}
+ 2 e^{2\frac{\phi_1}{\sqrt{6}}}) - e^{-2g} (a
e^{\frac{\phi_1}{\sqrt{6}}+\frac{\phi_2}{\sqrt{2}}} + 
b e^{\frac{\phi_1}{\sqrt{6}}-\frac{\phi_2}{\sqrt{2}}} - 
2 c e^{-2 \frac{\phi_1}{\sqrt{6}}} )}

\eqn{}{
\phi_2' = \frac{1}{\sqrt{2}}e^{f} [2 (- e^{-\frac{\phi_1}{\sqrt{6}}
-\frac{\phi_2}{\sqrt{2}}}  + 
e^{-\frac{\phi_1}{\sqrt{6}}+\frac{\phi_2}{\sqrt{2}}}) -  e^{-2g} (a
e^{\frac{\phi_1}{\sqrt{6}}+\frac{\phi_2}{\sqrt{2}}} - 
b e^{\frac{\phi_1}{\sqrt{6}}-\frac{\phi_2}{\sqrt{2}}} )}

A constant solution to these equations, can be
obtained iff $a=b=\half$ and reads, after
redefining 
$\varphi_1 =\frac{\phi_1}{\sqrt{6}}$ 
and 
$\varphi_2 =\frac{\phi_2}{\sqrt{2}}$ 
\eqn{}{f[r] = - Log[e^{2 \varphi_1} r],~~~~ \;\;\; e^{2g} =
\frac{e^{2\varphi_1}}{4}, ~~~~ \;\;\; 2 \cosh \varphi_2 =
e^{3\varphi_1}
}

We end this subsection by noting that according to \cite{Cvetic:1999xp}
the $g_{00}^{(10)}$ component of the 
ten dimensional metric is related to the  five dimensional $g_{00}^{(5)}$
 metric appearing in \lagrfd\ through 
\eqn{warp5}{g_{00}^{(10)} = W g^{(5)}_{00} ~,~~~~~~
W^2 = \left( e^{ - {\phi_1\over \sqrt{6} } - {\phi_2 \over \sqrt{2}} }
\sin^2\psi +e^{ - {\phi_1\over \sqrt{6} } + {\phi_2 \over \sqrt{2}} }
\cos^2\psi \right)\cos^2\theta +
 e^{ 2 {\phi_1\over \sqrt{6} }  }\sin^2\theta ~. }

\subsection{Effective Potential}
Here we consider the effective action for the configurations studied
above. Start by considering the Action

\eqn{}{
S=\int d^5 x \sqrt{G}[R - \frac{1}{2}\partial\phi^2 - V -\frac{1}{4}\Sigma
e^{2\alpha_i} F_{\mu\nu,i}^2 ] }
with  $\alpha_i$ as in \fieldsdef\ 
and 
\eqn{potencial}{V=- 4 \sum_{i}^{3}e^{\alpha_i}.}
Now consider a compactification to three dimensions on the space
\eqn{}{
ds_5^2 = ds_3^2 (r,z,t) + \frac{e^{2g}}{y^2}(dx^2 + dy^2)\;\;\; ~~~~
F_{x,y,i}=c_i/y^2 ,~~~~~~~\phi_2=0}
where we are thinking that we will quotient the $H^2$ space to obtain
a finite volume region. 
We get 
\eqn{}{
S \sim \int d^3 x \sqrt{g_3}e^{2g}[R(g_3) -
\frac{1}{2}g^{rr}(\partial_r\phi)^2 - 2 e^{-2 g}- V -\frac{1}{2}\Sigma
e^{2\alpha_i}c_i^2 e^{-4 g} ] ~. }
Transforming  to the Einstein frame
\eqn{einstfr}{
 g_3 = e^{- 4 g}  \hat{g_3}
}
we find 
\eqn{}{
S=\int d^3 x \sqrt{\hat{g_3}}[R(\hat{g_3}) -
\frac{1}{2}g^{rr}(\partial_r\phi)^2 - V_{eff} ]}
where,
\eqn{}{
 V_{eff}=e^{-4g} (\pm 2 e^{-2 g}  + V
+\frac{e^{-4g}}{2}(e^{2\frac{\phi}{\sqrt{6}}}a^2
+e^{2\frac{\phi}{\sqrt{6}}}b^2 +
e^{-4\frac{\phi}{\sqrt{6}}}c^2) )}
where $+$ is for genus  $g>1$ and $-$ is  for $S^2$.



\subsection{D=7}

Again in this case it is is convenient to choose a truncation of
$SO(5)$ gauged supergravity to a gauged  supergravity containing
only two $U(1)$ gauge fields. These two $U(1)$ fields correspond
to rotations in the 12 and 34 planes, where 1-5 are directions orthogonal
to the M5 brane.  We use \cite{Liu:1999ai} for the supersymmetry 
transformations and \cite{Cvetic:1999xp} for lifting up the solutions to 
eleven dimensions. 
 The effective Lagrangian is \cite{Liu:1999ai}
\ba
& &L= R- 5\partial(\lambda_{1}+
\lambda_{2}
)^2 - \partial(\lambda_{1}
-\lambda_{2}
)^2 - 
e^{-4\lambda_{1}
}
F_{\mu\nu,1}^2
-e^{-4\lambda_{2}
}
F_{\mu\nu,2}^2\nonumber\\
& &
-\frac{1}{2}m^2 (-8 e^{2(\lambda_{1}
+\lambda_{2}
)} - 4 e^{-2\lambda_{1}-4\lambda_{2}} - 4
e^{-4\lambda_{1}
-2\lambda_{2}}
+ e^{-8(\lambda_{1}+\lambda_{2})}) + L[C_3=0]
\ea
Again we will work in units where $R_{AdS_7} =1$ for the 
usual M5 solution. In these units the radius of $S^4$ is 1/2 and
$m=2$ above.

The supersymmetry transformations can be found in  \cite{Liu:1999ai},
\ba
& &\delta\psi_\mu = [\nabla_\mu + k/2 (A_\mu^{(1)} \Gamma^{12}
+A_\mu^{(2)} \Gamma^{34} ) + \frac{m}{4}e^{-4(\lambda_{1}
+\lambda_{2})}\gamma_\mu +
\frac{1}{2}\gamma_\mu\gamma^\nu\partial_\nu (\lambda_{1}
+\lambda_{2}
) +\nonumber\\
& &
\frac{1}{2}\gamma^\nu(e^{-2\lambda_{1}} F_{\mu\nu,1}\Gamma^{12}
+e^{-2\lambda_{2}} F_{\mu\nu,2}\Gamma^{34})]\epsilon
\ea
\eqn{trans2}{
\delta\lambda^{(1)} = [\frac{m}{4} (e^{2\lambda_{1}} - e^{-4(\lambda_{1}
+\lambda_{2})})
-\frac{1}{4}\gamma^\mu \partial_\mu (3\lambda_{1}
 + 2\lambda_{2}
)
-\frac{1}{8}\gamma^{\mu\nu}e^{-2\lambda_{1}}
F_{\mu\nu,1}\Gamma^{12}]\epsilon}
\eqn{trans3}{
\delta\lambda^{(2)} = [\frac{m}{4}(e^{2\lambda_{2}} - e^{-4(\lambda_{1}
+\lambda_{2})})
-\frac{1}{4}\gamma^\mu \partial_\mu (2\lambda_{1}
+ 3\lambda_{2})
-\frac{1}{8}\gamma^{\mu\nu}e^{-2\lambda_{2}}
F_{\mu\nu,2}\Gamma^{34}]\epsilon}

Where the spin connection components, for the metric
considered in the
text \metricse\ are given by
\eqn{spin7}{\omega_{\hat{u}}^{u,r}=  f'\;\; ~~~~ \omega_{\hat{x}}^{xr}=
\omega_{\hat{y}}^{yr}=
\frac{e^{g-f}g'}{y}\;\; ~~~~~~  \omega_{\hat{x}}^{xy}=
-\frac{1}{y}}
Here the hatted indices are curved while the unhatted one are flat.

The vanishing gravitino and gaugino equations  read 
\eqn{deltaspinor}{
\partial_{\hat{u}}\epsilon+\frac{\omega_{\hat{u}}^{u,r}\gamma_{ur}}{2}\epsilon
+
\frac{m}{4}
e^{-4(\lambda_{1}+\lambda_{2}) + f}\gamma_{u}\epsilon +
\frac{\lambda_{1}'+\lambda_{2}'}{2}\gamma_{ur}\epsilon=0}
\ba
& & [\partial_{\hat{x}}\epsilon +
\frac{1}{2}\omega_{\hat{x}}^{xy}\gamma_{xy}\epsilon
 + \frac{k}{2}(A_{\hat{x}}^{(1)} \Gamma^{12} +A_{\hat{x}}^{(2)}
\Gamma^{34})\epsilon +\frac{1}{2}\omega_{\hat{x}}^{xr}\gamma_{xr}\epsilon 
+ \frac{m}{4y}e^{-4(\lambda_1 +\lambda_2) + g}\gamma_x\epsilon\nonumber\\ 
& &+\frac{e^{g-f} (\lambda_1' +\lambda_2')}{2 y}\gamma_{xr}
+ \frac{y e^{-g}}{2} (e^{-2 \lambda_1}F_{xy}^{(1)} +
e^{-2\lambda_2}F_{xy}^{(2)})\gamma_y\epsilon=0
\ea
\eqn{}{m (e^{2\lambda_1} - e^{-4 (\lambda_1
+\lambda_2)})\epsilon - e^{-f} (3 \lambda_1' + 2 \lambda_2')
\gamma_r\epsilon - e^{-2g - 2\lambda_1} y^2 F_{xy}^{(1)} \gamma_{xy}
\Gamma^{(12)}\epsilon =0 }
\eqn{}{m (e^{2\lambda_1} - e^{-4 (\lambda_1
+\lambda_2)})\epsilon - e^{-f} (2 \lambda_1' + 3 \lambda_2')
\gamma_r\epsilon - e^{-2g - 2\lambda_2} y^2 F_{xy}^{(2)} \gamma_{xy}
\Gamma^{(34)}\epsilon =0 }
\eqn{}{\partial_{\hat{r}}\epsilon+
+ \frac{m}{4}
e^{-4(\lambda_{1}+\lambda_{2}) + f}\gamma_{r}\epsilon +
\frac{\lambda_{1}'+\lambda_{2}'}{2}\epsilon=0}

Imposing the conditions on the spinor
\eqn{spincondi}{\gamma_r\epsilon= \epsilon, \;\;, \gamma_{xy}\epsilon=
i \epsilon, \;\;\Gamma^{(12)}\epsilon = i
\epsilon,\;\;\Gamma^{(34)}\epsilon = i \epsilon, \;\;\;
\partial_{u,y,t,z,x,w}\epsilon=0,} 
Different choices of signs in \spincondi\ are related by simple
parity transformations.
we choose $\eta=\beta=\alpha=\gamma=1$ 
and evaluating over the configurations 
$A_{x}^{(1)} = \frac{a}{y}\;\;\;
A_{x}^{(2)} = \frac{b}{y}$

\eqn{susytransform}{\eqalign{
f'=& - e^{f -4(\lambda_1 +\lambda_2) } 
- \lambda_1' -\lambda_2' 
\cr
g' =& - e^{f} \left[\frac{1}{2}e^{-4 (\lambda_1 +\lambda_2)}
-  \kappa  e^{-2g} (a e^{-2 \lambda_1} +
b e^{-2\lambda_2}) \right] - (\lambda_1' + \lambda_2')
\cr
3\lambda_1' + 2\lambda_2' = & e^f \left[ 2 
(e^{2\lambda_1} - e^{-4 (\lambda_1+\lambda_2)}) + \kappa  a e^{-2g - 2
\lambda_1} \right]
\cr
3\lambda_2' + 2\lambda_1'  =& e^f \left[ 2 
(e^{2\lambda_2} - e^{-4 (\lambda_1+\lambda_2)}) + 
 \kappa b e^{-2g - 2 \lambda_2 } \right]
\cr
1 = & 4 (a + b)
}}
where $\kappa =1$ for $H^2$ and $\kappa = -1$ for $S^2$.
In the ${\cal N} =2$ configurations we set $a=1/4$, $b=0$,  
$3 \lambda_2 + 2 \lambda_1 =0$ and we define $\lambda = \lambda_2$.
Then we get the equations \gsev .
For ${\cal N} =1$ configurations we choose $a=b=1/8$ and 
$\lambda_1=\lambda_2 = -\phi/2 $ we get the equations \fseve .

We can also calculate the supergravity ``central charge'' of the 
four dimensional conformal field theory, which is given by
\eqn{centr}{
c = { \pi R^3_{AdS_5} \over 8  G_N^5}
}
in a normalization where $c = N^2/4$ for ${\cal N} =4 $ $U(N)$ SYM. 
In our case, a calculation parallel to that done for the D3 brane
gives 
\eqn{centrfive}{
c = { 8 \over 3}(g-1) e^{2 g + 3 f_0} N^3
}
where $f_0$ is the constant piece in $f$ defined as $f = f_0 - \log r $
where $f, ~r$ are those of the ansatz \metricse . 
{}From \fixedval\ and  \solusev\ we can see that we get $e^{2g + 3 f_0} =
1/8$
for the 
${\cal N} =2$ case and $e^{2g + 3 f_0} = 27/2^8$ for the ${\cal N} =1$ case.
In the ${\cal N} = 2$ case we can calculate the effective number of vector 
multiplets as follows. We know  that the conformal anomaly 
coefficients $a,c$ (see \cite{Aharony:1999ti,Henningson:1998gx}) 
are equal to leading order in $N$ for
theories
that have a gravity dual. This implies that this theory has the same number
of vector and hypermultiplets. Then from the formulas for $a,c$   in 
\cite{Duff:1994wm} and \centr\  we find  $n_V = n_H = { 4 \over 3} (g-1)
N^3$.

Let us finish this appendix by writing  the expression for 
the warp factor that is
obtained when uplifting this solutions to M theory \cite{Cvetic:1999xp}
\eqn{warp7}{ g_{00}^{(11)} = W g_{00}^{(7)} ~,~~~~~~~
W^3= e^{-4(\lambda_1+\lambda_2)} \mu_0^2 +   e^{2 \lambda_2}\mu_1^2 +
e^{2 \lambda_1} \mu_2^2}
where $\mu_i$ parametrize $S^2$, $\mu_0^2+\mu_1^2 + \mu_2^2=1$.

\subsection{D4 brane solutions}

Here we consider the solutions that we get when we compactify a direction
along the M5 worldvolume on $S^1$ and we reduce the 11 dimensional solution
to  a 10 dimensional solution describing a D4 brane wrapped on the
Riemann surface.

We find that the dilaton and ten dimensional metric are
\eqn{d4sol}{\eqalign{
e^{2 \phi} = & W e^{ 3 f}
\cr
ds^2_{str} =& e^{ 2 \phi/3} \left. ds^2_{11}\right|_{10}
\cr
W^2  = & 
e^{ 2 \lambda } \cos^2 \theta  + e^{-3 \lambda} \sin^2 \theta 
}}
where the right hand side of the second line denotes the eleven 
dimensional metric along the ten directions of the type IIA solution,
the factor of the dilaton produces the string metric in ten dimensions.
The functions $\lambda, ~f$ that appear here are those of the 
solution \solmfi .
In principle we could also find the various components of the 
four form field strength of type IIA from that of the 11 dimensional
solution. Of course we can do the same with the other solution we
found, the one with ${\cal N} =1$ supersymmetry in four dimensions.

\renewcommand{\baselinestretch}{0.87}
\footnotesize

\bibliography{paper}
\bibliographystyle{ssg}

\end{document}